\documentclass[12pt,a4paper]{article}
\usepackage{graphicx}
\usepackage{subeqnarray}
\begin{document}
%
%
%
%
\def\astrobj#1{#1}
\newenvironment{lefteqnarray}{\arraycolsep=0pt\begin{eqnarray}}
{\end{eqnarray}\protect\aftergroup\ignorespaces}
\newenvironment{lefteqnarray*}{\arraycolsep=0pt\begin{eqnarray*}}
{\end{eqnarray*}\protect\aftergroup\ignorespaces}
\newenvironment{leftsubeqnarray}{\arraycolsep=0pt\begin{subeqnarray}}
{\end{subeqnarray}\protect\aftergroup\ignorespaces}
\newcommand{\diff}{{\rm\,d}}
\newcommand{\pprime}{{\prime\prime}}
\newcommand{\szeta}{\mskip 3mu /\mskip-10mu \zeta}
\newcommand{\FC}{\mskip 0mu {\rm F}\mskip-10mu{\rm C}}
\newcommand{\appleq}{\stackrel{<}{\sim}}
\newcommand{\appgeq}{\stackrel{>}{\sim}}
\newcommand{\Int}{\mathop{\rm Int}\nolimits}
\newcommand{\Nint}{\mathop{\rm Nint}\nolimits}
\newcommand{\range}{{\rm -}}
\newcommand{\displayfrac}[2]{\frac{\displaystyle #1}{\displaystyle #2}}
\def\astrobj#1{#1}
%
\title{Evolving statistical systems: 
\\ application to academic courses}

\author{ 
{R.~Caimmi}\footnote{
{\it Physics and Astronomy Department, Padua University,
Vicolo Osservatorio 3/2, I-35122 Padova, Italy.  Affiliated up to September
30th 2014.   Current status: Studioso Senior.  Current position: in retirement
due to age limits.}\hspace{50mm}
email: roberto.caimmi@unipd.it~~~
fax: 39-049-8278212}
\phantom{agga}}
%
%
\maketitle
\begin{quotation}
\section*{}
\begin{Large}
\begin{center}

Abstract

\end{center}
\end{Large}
\begin{small}

\noindent\noindent
Statistical systems are conceived from the standpoint of statistical
mechanics, as made of a (generally large) number of identical units and
exhibiting a (generally large) number of different configurations
(microstates), among which only equivalence classes (macrostates) are
accessible to observations.   Further attention is devoted to evolving
statistical systems, and a simple case including only a possible event, E, and
related opposite event, $\neg$E, is examined in detail.   In particular, the
expected evolution is determined and compared to the random evolution inferred
from a sequence of random numbers, for different sample populations.   The
special case of radioactive decay is considered and results are expressed in
terms of the fractional time, $t/\Delta t$, where the time step, $\Delta t$,
is related to the decay probability, $p=p(\Delta t)$.   An application is made
to data collections from selected academic courses,
focusing on the extent to which expected evolutions and model random
evolutions fit to empirical random evolutions inferred from data collections.
Results could be biased by the assumed number of students who abandoned their
course, defined as suitable impostors (SI).   Extreme cases related to a lower
and an upper limit of the SI number are considered for a time step,
$\Delta t=(1/12)$y, where fitting expected evolutions relate to
$0.003\le p\le0.200$.   In conclusion, evolving statistical systems made of
academic courses are similar to poorly populated samples of radioactive
nuclides
exhibiting equal probabilities, $p$, and time steps, $\Delta t$, where
inferred mean lifetimes, $\tau$, and half-life times, $t_{1/2}$, range within
$0.37<\tau/{\rm y}<27.73$ and $0.25<t_{1/2}/{\rm y}<19.22$, respectively, and
upper limits are related to incomplete data collections.

\noindent
{\bf Keywords:} Systems: statistical; events: microstates, macrostates;
evolutions: expected, random.
\end{small}
\end{quotation}

\section{Introduction} \label{intro}

From the standpoint of statistical mechanics, statistical systems can be
conceived as made of a (in general, extremely large) number of identical
``atomic'' units and exhibiting a (in general, extremely large) number of
distinct configurations (microstates), among which only equivalence classes
(macrostates) are accessible to observations.   More specifically, a single
microstate arises after a selected physical process, or attempt, is applied
to each unit, which can attain two distinct configurations related to a
possible event, E, and the opposite event, $\neg$E, respectively.
Accordingly, microstates can be defined as sequences of E and $\neg$E, or
$n$-tuples if the system under consideration is made of $n$ units, and
macrostates can be defined as subsets of $n$-tuples exhibiting $k$ possible
events, E, and $(n-k)$ opposite events, $\neg$E, $0\le k\le n$.

For instance, a unit may be a box containing $N$ identical (leaving aside the
colour) spheres, numbered on the inside in arithmetic progression, among which
$N_{\rm W}$ are white and $(N-N_{\rm W})$ are black.   Accordingly, the
attempt is the extraction of a sphere; the possible event, E, is a white
sphere extracted, and the opposite event, $\neg$E, is a black sphere
extracted.

If the system is made of $n$ boxes, numbered on the inside in
arithmetic progression, a microstate can be denoted by a $n$-tuple,
$\{i_1,i_2,...,i_n\}$, where $i_k$ denotes the $i$th sphere within the $k$th
box.

If numbered spheres and numbered boxes are not accessible to
observations, macrostates are denoted by $n$-tuples of events (or colours),
\linebreak
$\{\{\{{\rm E}^k\neg{\rm E^{n-k}}\}\}\}$, where inner brackets represent a
single microstate made of $k$ specified whithe spheres and $(n-k)$ specified
black spheres, each extracted from $k$ specified boxes and $(n-k)$ specified
boxes, respectively; middle brackets represent the whole set of microstates
made of $k$ unspecified whithe spheres and $(n-k)$ unspecified
black spheres, each extracted from $k$ specified boxes and $(n-k)$ specified
boxes, respectively; outer brackets represent the whole set of microstates
made of $k$ unspecified whithe spheres and $(n-k)$ unspecified black spheres,
each extracted from $k$ unspecified boxes and $(n-k)$ unspecified boxes,
respectively.

The probability of macrostates is expressed by the binomial distribution, as
$P_n(k)={n\choose k}p^kq^{n-k}$, where $p$ is the probability of the possible
event, E, and $q=1-p$ is the probability of the opposite event, $\neg$E, made
of a white and a black sphere extracted, respectively, in the above mentioned
example.   Related expectation value and variance are $k^\ast=np$ and
$\sigma_{\rm B}=npq$, respectively.

Statistical systems where attempts are repeated at a fixed frequency,
$(\Delta t)^{-1}$, can either remain unchanged or evolve.   For instance,
boxes where extracted spheres are reintroduced before the next extraction 
remain unchanged, while boxes are evolving if the contrary holds.   A special
class of evolving statistical systems relates to units which are maintained or
removed according if the possible event, E, or the opposite event, $\neg$E,
respectively, occurs after an attempt is performed.   For instance, the
extraction of white or black spheres could imply box preservation or removal,
respectively.   Typical evolving statistical systems are (i) samples of
radioactive nuclides, and (ii) selected academic courses.

The current paper is aimed to highlight some aspects of evolving statistical
systems with regard to the fractional number of surviving units, $n(t)/n_0$,
where $n_0$ is the initial number.   In particular, attention is devoted to
radioactive decay and an application is made to data collection related to
selected academic courses.

Basic considerations on statistical systems, in the light of statistical
mechanics, are outlined in Section \ref{stas}.   Evolving statistical systems
are considered in Section \ref{ests}, where further attention is devoted to
radioactive decay.   An application to data collections related to selected
academic courses is performed in Section \ref{aaco}.   The discussion is
presented in Section \ref{disc}.  The conclusion is drawn in Section
\ref{conc}.   Further details on data collection and input parameters are
shown in the Appendix.

\section{Statistical systems} \label{stas}
\subsection{General remarks} \label{gere}

\noindent\noindent
Statistical systems can be defined as able to attain a number of different
configurations after experiencing specified physical or conceptual processes
in ordinary or abstract space, respectively.   Configurations attained by
statistical systems can be defined as possible events, and the whole amount of
selected physical or conceptual processes can be defined as attempt.

Let S$_1$ be a simple statistical system i.e. made of a single unit.   Let the
effect of an attempt, A$_1$, performed on S$_1$, be the possible event, E$_1$,
or the opposite event, $\neg$E$_1$, i.e. the union of all possible events
other than E$_1$.   Let S$_{u,1}$ be an underlying statistical system which
exhibits the following features.
\begin{description}
\item[(i)\hspace{2.7mm}] E$_1$ occurs in S$_1$ if and only if (E$_{u,1})_i$,
$1\le i\le N_u({\rm E}_{u,1})$, occurs in S$_{u,1}$, where
$N_u({\rm E}_{u,1})$ is
the number of possible occurrences of E$_{u,1}$ in S$_{u,1}$;
\item[(ii)~] $\neg$E$_1$ occurs in S$_1$ if and only if ($\neg$E$_{u,1})_j$,
$1\le j\le N_u(\neg{\rm E}_{u,1})$, occurs in S$_{u,1}$, where
$N_u(\neg{\rm E}_{u,1})$ is the
number of possible occurrences of $\neg$E$_{u,1}$ in S$_{u,1}$;
\item[(iii)\hspace{.7mm}] There is no physical or conceptual reason for which,
after performing an attempt, A$_{u,1}$, on S$_{u,1}$, (E$_{u,1})_i$ has to be
preferred with respect to ($\neg$E$_{u,1})_j$ or (E$_{u,1})_j$, and
($\neg$E$_{u,1})_j$ has to be preferred with respect to ($\neg$E$_{u,1})_i$ or
(E$_{u,1})_i$.
\end{description}
Accordingly, the probability of the possible event, E$_1$, can be defined as
$p=N_u({\rm E}_{u,1})/N_u$ and the probability of the opposite event,
$\neg$E$_1$,
can be defined as $q=N_u(\neg{\rm E}_{u,1})/N_u$, where
$N_u=N_u({\rm E}_{u,1})+N_u(\neg{\rm E}_{u,1})$, which implies the
normalization condition, $p+q=1$.

In the light of quantum mechanics, physical processes involve discrete
quantities, which implies $p$ is a rational number.   On the other hand e.g.,
in the light of geometry, conceptual processes could involve continuous
quantities, which implies $p$ could be an irrational number.   If it is the
case, both
$N_u({\rm E}_{u,1})$ and $N_u(\neg{\rm E}_{u,1})$ must be infinite to ensure a
ratio, $N_u({\rm E}_{u,1})/N_u$ and $N_u(\neg{\rm E}_{u,1})/N_u$, infinitely
close to $p$ and $q$, respectively, via Dedekind's axiom.

Let S$_n$ be a complex statistical system made of $n$ units S$_1$.   Let A$_n$
be an attempt performed on S$_n$, made of $n$ attempts A$_1$ each performed on
a different S$_1$.   Let
$s_{u,k}={\rm E}_{u,k}=\{{\rm E}_{u,1}^k\neg{\rm E}_{u,1}^{n-k}\}$,
$0\le k\le n$, denote a configuration of S$_n$ after A$_n$ has been performed.
More specifically, E$_{u,k}$ is a complex event resulting from the union of
$k$ independent events, (E$_{u,1})_i$, $1\le i\le N_u$, each related to a
specified S$_n$, and $(n-k)$ independent events, $\neg({\rm E}_{u,1})_j$,
$1\le j\le N_u$, each related to a specified S$_n$.

Similarly, $s_{u,k}^
\prime=\{{\rm E}_{u,k}\}=\{\{{\rm E}_{u,1}^k\neg{\rm E}_{u,1}^{n-k}\}\}$,
$0\le k\le n$, denotes the whole set of ${\rm E}_{u,k}$, in number of
$[N_u({\rm E}_{u,1}]^k[N_u-N_u({\rm E}_{u,1})]^{n-k}$, where $k$ independent
events, E$_{u,1}$, and $(n-k)$ independent events, $\neg$E$_{u,1}$, are no
longer specified.

Finally,
$s_{n,k}=\{\{\{{\rm E}_{u,1}^k\neg{\rm E}_{u,1}^{n-k}\}\}\}=\{\{{\rm E}_1^k
\neg{\rm E}_1^{n-k}\}\}=\{{\rm E}_{1,k}\}$, $0\le k\le n$, denotes the whole
set of ${\rm E}_{1,k}$, in number of ${n\choose k}$, where $k$ units, S$_1$,
related to E$_1$, and $(n-k)$ units, S$_1$, related to $\neg$E$_1$, are no
longer specified.

In short, $s_{u,k}$ can be conceived as a microstate and $s_{n,k}$ as a
macrostate, with regard to an attempt, A$_n$, performed on a complex
statistical system, S$_n$.   The whole set of microstates yields the sure
event, E$_{{\rm S},n}$.   The empty set of microstates yields the unpossible
event, E$_{{\rm U},n}$.   In the case under discussion, the number of
microstates related to $s_{u,k}^\prime$, $s_{n,k}$, E$_{{\rm S},n}$,
E$_{{\rm U},n}$, reads:
\begin{leftsubeqnarray}
\slabel{eq:Na}
&& N(s_{u,k}^\prime)=[N_u({\rm E}_{u,1}]^k[N_u-N_u({\rm E}_{u,1})]^{n-k}~~; \\
\slabel{eq:Nb}
&& N(s_{n,k})={n\choose k}[N_u({\rm E}_{u,1}]^k[N_u-N_u({\rm E}_{u,1})]^{n-k}
~~; \\
\slabel{eq:Nc}
&& N({\rm E}_{{\rm S},n})=N_u^n~~;\quad N({\rm E}_{{\rm U},n})=0~~;
\label{seq:N}
\end{leftsubeqnarray}
and the corresponding probability is:
\begin{leftsubeqnarray}
\slabel{eq:Pa}
&& P(s_{u,k}^\prime)=\frac{N(s_{u,k}^\prime)}{N({\rm E}_{{\rm S},n})}=p^kq^
{n-k}~~; \\
\slabel{eq:Pb}
&& P(s_{n,k})={n\choose k}p^kq^{n-k}~~; \\
\slabel{eq:Pc}
&& P({\rm E}_{{\rm S},n})=\frac{N({\rm E}_{{\rm S},n})}
{N({\rm E}_{{\rm S},n})}=1~~;\quad P({\rm E}_{{\rm U},n})=\frac
{N({\rm E}_{{\rm U},n})}{N({\rm E}_{{\rm S},n})}=0~~;
\label{seq:P}
\end{leftsubeqnarray}
according to the above considerations.

Let the macrostate, $s_{n,k}$, be designed after performing an attempt, A$_n$,
on a complex statistical system, S$_n$.   The variable, $k$, $0\le k\le n$, is
a random variable, and the probability of a macrostate, $P_n(k)$, is the
related distribution.   In particular, the distribution expressed by
Eq.\,(\ref{eq:Pb}):
\begin{equation}
\label{eq:bino}
P_n(k)={n\choose k}p^kq^{n-k}~~;\quad0\le k\le n~~;
\end{equation}
 is known as binomial distribution or Bernoulli distribution.   By definition,
 related expectation value and variance read:
\begin{lefteqnarray}
\label{eq:exbi}
&& k^\ast=\sum_{k=0}^nkP_n(k)=np~~; \\
\label{eq:vabi}
&& \sigma_{\rm B}^2=\sum_{k=0}^n(k-k^\ast)^2P_n(k)=npq~~;
\end{lefteqnarray}
where the index, B, denotes binomial distribution.

In the limit, $n\to+\infty$, $p={\rm const}$, the binomial distribution takes
the expression:
\begin{equation}
\label{eq:gaus}
\lim_{n\to+\infty}P_n(k)=\lim_{n\to+\infty}\frac h{\sqrt\pi}
\exp[-h^2(k-k^\ast)^2]~~;\quad h=\frac1{2npq}~~;\quad p={\rm const}~~;
\end{equation}
which is known as Gauss distribution (with regard to a single source of
accidental errors).   Related expectation value and variance are divergent via
Eqs.\,(\ref{eq:exbi}) and (\ref{eq:vabi}).

In  the limit, $n\to+\infty$, $np=$ const, the binomial distribution takes the
expression:
\begin{equation}
\label{eq:pois}
\lim_{n\to+\infty}P_n(k)=\frac{(k^\ast)^k}{k!}\exp(-k^\ast)~~;
\quad k^\ast=np={\rm const}~~;
\end{equation}
which is known as Poisson distribution.   Related expectation value and
variance are expressed via Eqs.\,(\ref{eq:exbi}) and (\ref{eq:vabi}), the last
reduced to:
\begin{equation}
\label{eq:vapo}
\sigma_{\rm P}^2=np=k^\ast~~;
\end{equation}
where the index, P, denotes Poisson distribution.
For further details, an interested reader is addressed to specific textbooks
e.g., \cite{cai16} Chap.\,2.

\subsection{A guidance example} \label{guex}

The following guidance example is aimed to better understanding considerations
outlined in Subsection \ref{gere}.   In this view, S$_{u,1}$ is conceived as a
box, B$_1$, containing $N$ identical spheres numbered on the inside in
arithmetic progression $(i=1,2,...,N)$, among which $N_{\rm W}$ are white and
$N-N_{\rm W}$ are black.   The possible event, E$_{u,1}$, is a white sphere
extracted from B$_1$; the opposite event, $\neg$E$_1$, is a black sphere
extracted from B$_1$; a single extraction is the related attempt.

Requirements mentioned in Subsection \ref{gere} are satisfied, as (i) E$_1$ is
the union of $N_{\rm W}$ white spheres which can be extracted from B$_1$,
$({\rm E}_{u,1})_i$, $1\le i\le N_{\rm B}$; (ii) $\neg$E$_1$ is the union of
$(N-N_{\rm W})$ black spheres which can be extracted from B$_1$,
$(\neg{\rm E}_{u,1})_i$, $1\le i\le N-N_{\rm B}$; (iii) there is no physical
or conceptual reason for which, after performing an extraction from B$_1$, a
specified sphere has to be preferred with respect to another one.
Accordingly, the probability of extracting a specified (white or black) sphere
is $p_i=q_i=1/N$, and the probability of extracting an unspecified white or
black sphere is $p=N_{\rm W}/N$ or $q=(N-N_{\rm W})/N$, respectively.

The possible event, E$_1$, is statistically equivalent to a white sphere
extracted from B$_1$ and the opposite event, $\neg$E$_1$, is statistically
equivalent to a black sphere extracted from B$_1$, in the sense that
$p({\rm E}_1)=p=N_{\rm W}/N$ and $p(\neg{\rm E}_1)=q=(N-N_{\rm W})/N$,
respectively.

With regard to the complex statistical system, B$_n$, made of $n$ units,
B$_1$, and to the attempt, A$_n$, made of $n$ attempts, A$_1$, performed each
on a different B$_1$, the microstate, $s_{u,k}$, is made of $k$ specified
white spheres and $(n-k)$ specified black spheres, each extracted from a
different specified box.

Similarly, $s_{u,k}^\prime$, is made of the whole set of $s_{u,k}$, where
there are $N_{\rm W}$ different white spheres and $(N-N_{\rm W})$ different
black spheres to be extracted from each specified box, for a total of
$N_{\rm W}^k(N-N_{\rm W})^{n-k}$.

Finally the macrostate, $s_{n,k}$, is made of the whole set of
$s_{u,k}^\prime$, where there are ${n\choose k}$ different ways of extracting
$k$ white spheres and $(n-k)$ black spheres from $n$ identical boxes.

For $n\gg1$, microstates are virtually indistinguishable (leaving aside
Laplace's daemon) and only macrostates can be detected.   In the case under
discussion, microstates are denoted by numbers specifying spheres and boxes,
and macrostates by colours.   Natural limits intrinsic to observers makes
numbers undetected and colours distinguishable.

With regard to a generic statistical system, S$_1$, where an attempt, A$_1$,
is performed
yielding either the possible event, E$_1$, or the opposite event, $\neg$E$_1$,
of probability, $P({\rm E}_1)=p$, $P(\neg{\rm E}_1)=q=1-p$, respectively, the
underlying statistical system, S$_{u1}$, may be conceived as a box, B$_1$,
containing $N$ identical spheres among which $N_{\rm W}$ are white and
$N-N_{\rm W}$ are black, where $p=N_{\rm W}/N$ and $q=(N-N_{\rm W})/N$,
respectively.   For further details, an interested reader is addressed to
specific textbooks e.g., \cite{cai16} Chap.\,2.
 
\section{Evolving statistical systems} \label{ests}
\subsection{Basic considerations} \label{baco}

With regard to a statistical system, S$_n$, made of $n$ units, S$_1$, and to
an attempt, A$_n$, made of the union of $n$ attempts, A$_1$, each in
connection with
a different S$_1$, performing A$_n$ on S$_n$ yields a macroscopical state,
$s_{n,k}$, $0\le k\le n$.   Related probability obeys binomial
distribution, as expressed by Eq.\,(\ref{eq:bino}).   If S$_n$ maintains
unchanged after undergoing a succession of A$_n$, no evolution occurs.   If
the contrary holds, some kind of evolution takes place and an arrow of the
time can be defined.

For instance, let S$_1$ be a box containing identical white and black spheres,
A$_1$ a single extraction, E$_1$ a white sphere extracted and $\neg$E$_1$ a
black sphere extracted.   Let A$_n$ be performed on S$_n$ and $s_{n,k_1}$ be
denoted.   Let the additional condition hold that all boxes, from which white
spheres are extracted, be removed before the next attempt.   Accordingly,
S$_n$, A$_n$, are changed into S$_{n-k_1}$, A$_{n-k_1}$, respectively, and so
on until $n-k_1-...-k_L=0$ after $L$ successive attempts.   Statistical
systems of the kind considered can be conceived as evolving.

Let S$_n$ be an evolving statistical system, and let a trial, A$_n$, be
successively performed on S$_n$ after a time step, $\Delta t$, has been
elapsed.   Let $s_{n_\ell, k_\ell}$ be a macrostate at the $\ell$th step.
Related binomial distribution via Eq.\,(\ref{eq:bino}) reads:
\begin{equation}
\label{eq:bine}
P_{n_\ell}(k_\ell)={n_\ell\choose k_\ell}p^{k_\ell}q^{n_\ell-k_\ell}~~;\quad
0\le k_\ell\le n_\ell~~;
\end{equation}
which exhibits expectation value and variance via Eqs.\,(\ref{eq:exbi}) and
(\ref{eq:vabi}), respectively, as:
\begin{lefteqnarray}
\label{eq:exbe}
&& k_\ell^\ast=n_{\ell-1}p~~; \\
\label{eq:vabe}
&& \sigma_{{\rm B},\ell}^2=n_{\ell-1}pq~~;
\end{lefteqnarray}
where $p=p(\Delta t)$ and $q=q(\Delta t)=1-p(\Delta t)$ is the probability of
E$_1$ and $\neg$E$_1$, respectively, after a time step, $\Delta t$, has been
elapsed.

\subsection{Expected evolution} \label{exev}

With regard to an evolving statistical system, S$_n$, and an attempt, A$_n$,
let $p=p(\Delta t)$ be the probability of the event, E$_1$, in connection with
a generic unit, S$_1$, within a time step, $\Delta t$, where the occurrence of
E$_1$ implies related S$_1$ is removed from S$_n$.

The expected number of surviving units,
at the end of the $\ell$th step, via Eq.\,(\ref{eq:exbe}) reads:
\begin{equation}
\label{eq:enla}
n_\ell^\ast=n_{\ell-1}^\ast-k_\ell^\ast=n_{\ell-1}^\ast(1-p)~~;\quad
n_0^\ast=n_0~~;
\end{equation}
which, after $\ell$ iterations, yields the fractional number of surviving
units as:
\begin{lefteqnarray}
\label{eq:enlr}
&& \frac{n_\ell^\ast}{n_0}=(1-p)^\ell~~; \\
\label{eq:l}
&& \ell=\frac{t_\ell-t_0}{\Delta t}~~;
\end{lefteqnarray}
where $t_\ell$ is the time at the end of the $\ell$th step and $t_0$ is the
initial time.   It is worth noticing $p$ depends on the time step only, while
$\ell$ depends on both the time elapsed and the time step.

Using the logarithmic Taylor series e.g., \cite{spi68} Chap.\,20 \S20.17:
\begin{equation}
\label{eq:lose}
\ln(1+x)=x-\frac{x^2}2+\frac{x^3}3-\frac{x^4}4+...~~;\quad-1<x\le1~~;
\end{equation}
the following identity holds:
\begin{equation}
\label{eq:epa1}
(1-p)^\ell=\exp\ln(1-p)^\ell=\exp[\ell\ln(1-p)]=\exp\left[-\ell\left(p+
\frac{p^2}2+\frac{p^3}3+\frac{p^4}4+...\right)\right];
\end{equation}
which, in the limit of negligible $p$, reduces to:
\begin{equation}
\label{eq:epa2}
(1-p)^\ell=\exp(-p\ell)~~;\quad p\ll1~~;
\end{equation}
and the substitution of Eq.\,(\ref{eq:epa2}) into (\ref{eq:enlr}) yields:
\begin{equation}
\label{eq:enre}
\frac{n_\ell^\ast}{n_0}=\exp\left(-p\frac{t_\ell-t_0}{\Delta t}\right)~~;
\end{equation}
where the product, $p\ell=p(\Delta t)(t_\ell-t_0)/\Delta t$, for fixed
$t_\ell$ remains unchanged provided the ratio, $p(\Delta t)/\Delta t$, remains
unchanged i.e. $p$ is directly proportional to $\Delta t$.

The ratio, $(1-p)^\ell/\exp(-p\ell)$, by use of Eq.\,(\ref{eq:epa1}) reads: 
\begin{lefteqnarray}
\label{eq:cfea}
&& \frac{(1-p)^\ell}{\exp(-p\ell)}=\exp\left[-\ell\left(p+\frac{p^2}2+\frac
{p^3}3+\frac{p^4}4+...\right)\right]\exp(p\ell) \nonumber \\
&& \phantom{\frac{(1-p)^\ell}{\exp(-p\ell)}}=
\exp\left[-\ell\left(\frac{p^2}2+\frac{p^3}3+\frac{p^4}4+...\right)\right]~~;
\end{lefteqnarray}
which tends to zero as $\ell\to+\infty$ i.e. an infinite time.   Accordingly,
the power, $(1-p)^\ell$, is overstimated by the exponential, $\exp(-p\ell)$,
and the former is infinitesimal of higher order with respect to the latter, as
$\ell\to+\infty$.

The above results hold for the expected evolution, where the number of
surviving S$_1$ at the end of any step equals related expected number.
According to Bernoulli's theorem, discrepancies between random evolution and
expected evolution may safely thought of as negligible for large S$_1$
populations, $n\gg1$.

\begin{figure*}[t]  
\begin{center}      
\includegraphics[scale=0.8]{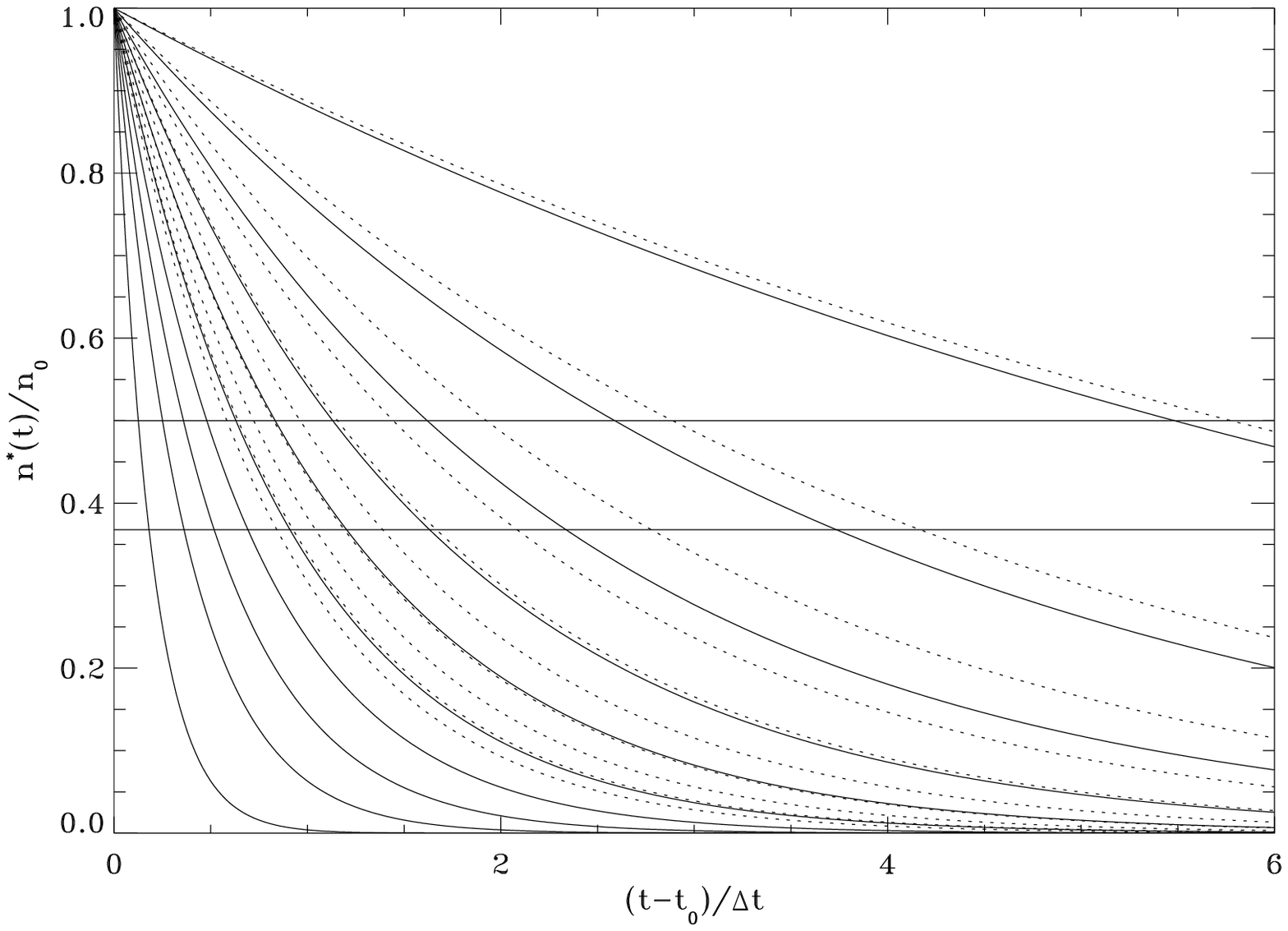}                      
\caption[ddbb]{Expected evolution of fractional number of surviving units,
$n^\ast(t)/n_0$, related to both exact power law (full curves) and exponential
approximation (dotted curves) for different values of probability, 
$p=p(\Delta t)=(i-\delta_{i,10}/10)/10$, $1\le i\le10$.   The limit, $p=0$,
corresponds to the top side of the box.   The limit, $p=1$, corresponds to the
left side of the box.   The fractional time,
$(t_\ell-t_0)/\Delta t$, corresponds to the number of steps, $\ell$.  The
changes, $(t-t_0)\to k(t-t_0)$, $\Delta t\to k\Delta t$, $k>0$, imply 
$p(\Delta t)\to p(k\Delta t)$ for any point.   The mean lifetime, $\tau$, and
the mean half-life time, $t_{1/2}$, can be inferred from the intersection of
the selected curve with the horizontal line,
$n^\ast(t)/n_0=1/{\rm e}\approx0.3679$
and $n^\ast(t)/n_0=1/2$, respectively.   See text for further details.}
\label{f:cute}     
\end{center}       
\end{figure*}

The expected evolution of fractional number of surviving units, $n_\ell^\ast$,
is shown in Fig.\,\ref{f:cute} for both the exact power law, expressed by
Eq.\,(\ref{eq:enlr}), and the exponential approximation, expressed by
Eq.\,(\ref{eq:enre}).
%
Different
curves relate to different probabilities, $p=P({\rm E}_1)$, within the range,
$0.1\le p\le 0.99$.   The limit, $p=0$, corresponds to $n^\ast(t)/n_0=1$.
The limit, $p=1$, corresponds to $(t-t_0)/\Delta t=0$.

A generic point on the $\{{\sf O}[(t-t_0)/\Delta t][n^\ast(t)/n_0]\}$ plane
depends on three parameters, namely
the time elapsed, $t-t_0$, the time step, $\Delta t$, and the probability,
$p=p(\Delta t)$.   The changes, $(t-t_0)\to k(t-t_0)$,
$\Delta t\to k\Delta t$, leave a generic point unchanged provided
$p(\Delta t)\to p(k\Delta t)$.    It is worth of note the expected evolution
needs an infinite time via Eq.\,(\ref{eq:enlr}) or (\ref{eq:enre}), while the
random evolution ends within a finite time unless $n_0\to+\infty$.

\subsection{Random evolution} \label{raev}

With regard to an evolving statistical system, S$_n$, and an attempt, A$_n$,
let $p=p(\Delta t)$ be the probability of the event, E$_1$, within a time
step, $\Delta t$, where the occurrence of E$_1$ on a generic unit, S$_1$,
implies removal from S$_n$.   Accordingly, the random evolution can be
determined along the following steps e.g., \cite{wk95} \cite{cai00}.
\begin{description}
\item[(i)\hspace{2.9mm}] Take $\ell=0$ at the beginning of the first step.
\item[(ii)\hspace{2.7mm}] Generate a succession of $n_\ell$ random numbers,
$\xi_1, \xi_2, ..., \xi_{n_\ell}$, within the range, $0\le\xi_i\le1$,
$1\le i\le n_\ell$.
\item[(iii)~]Perform a one-to-one correspondance,
$({\rm S}_1)_i\leftrightarrow\xi_i$, $1\le i\le n_\ell$.
\item[(iv)\hspace{.7mm}]Remove $({\rm S}_1)_i$ unit from S$_{n_\ell}$ if
$\xi_i<p$ and preserve if otherwise, yielding
S$_{n_\ell}(n_\ell)\to$S$_{n_\ell-\Delta n_\ell}(n_\ell-\Delta n_\ell)=
$S$_{n_{\ell+1}}(n_{\ell+1})$, where $\Delta n_\ell$
is the number of removed units at the end of the $(\ell+1)$th step.
\item[(v)\hspace{1mm}]End if $n_{\ell+1}=0$ or replace $n_\ell$ with
$n_{\ell+1}$ and return to (ii) if $n_{\ell+1}>0$.
\end{description}
Owing to finite $n_0$, the random evolution ends when, after $L$ steps,
$n_L=0$ and $S_{n_L}=S_0$ is the empty set of units.

\begin{figure*}[t]  
\begin{center}      
\includegraphics[scale=0.8]{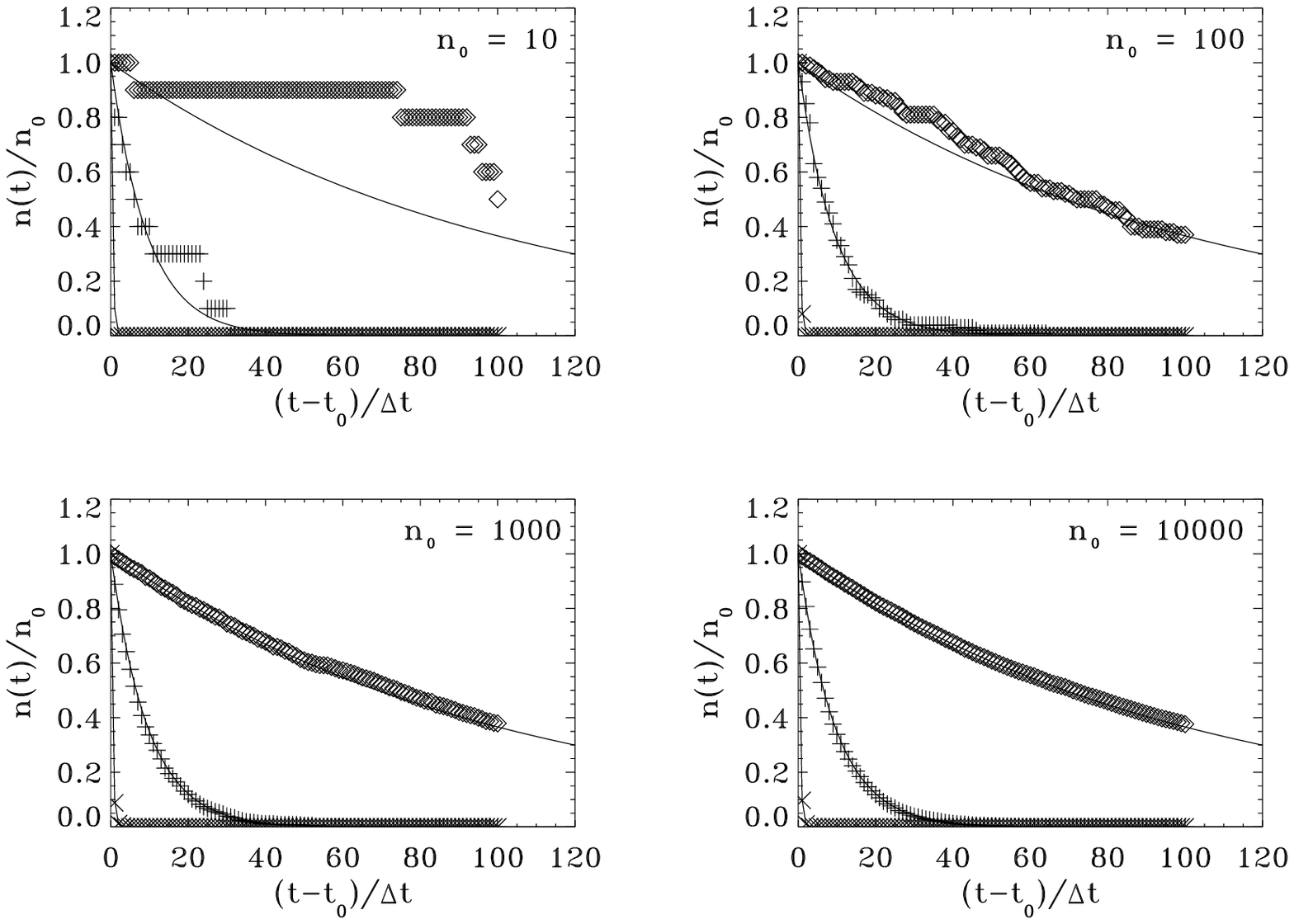}                      
\caption[ddbb]{Example of model random evolution of fractional number of
surviving
units, $n(t)/n_0$, for $\log n_0=1,2,3,4$, listed on each panel, and $p=0.01$
(diamonds), 0.1 (crosses), 0.9 (saltires).   Related expected evolution is
shown by full lines. The fractional time, $(t_\ell-t_0)/\Delta t$, corresponds
to the number of steps, $\ell$.   See text for further details.}
\label{f:nebi}     
\end{center}       
\end{figure*}

An example of model random evolution is shown in Fig.\,\ref{f:nebi} for
$\log n_0=1,2,3,4$ (listed on each panel) and $p=0.01$ (diamonds), 0.10
(crosses), 0.90 (saltires).   
Related expected evolution is represented as full curves.   An inspection of
Fig.\,\ref{f:nebi} discloses statistical fluctuations are large for
$n_0\appleq10$, small for $10\ll n_0\appleq100$, negligible for
$100\ll n_0\appleq1000$ and $n_0>1000$.

\subsection{Mean lifetime and half-life time} \label{taut}

With regard to an evolving statistical system, S$_n$, where the exponential
approximation holds to an acceptable extent, let the mean lifetime, $\tau$,
and the mean half-life time, $t_{1/2}$, be defined in connection with the
following values of the fractional number of surviving units:
\begin{lefteqnarray}
\label{eq:nata}
&& \frac{n^\ast(\tau)}{n_0}=\exp(-1)~~; \\
\label{eq:nati}
&& \frac{n^\ast(t_{1/2})}{n_0}=\frac12~~;
\end{lefteqnarray}
where Eq.\,(\ref{eq:nata}) via (\ref{eq:enre}) implies an explicit expression
of the mean lifetime as:
\begin{lefteqnarray}
\label{eq:taua}
&& \tau=\frac{\Delta t}p~~;\quad p\ll1~~;
\end{lefteqnarray}
accordingly, Eq.\,(\ref{eq:enre}) reads:
\begin{lefteqnarray}
\label{eq:ntaa}
&& \frac{n_\ell^\ast}{n_0}=\exp\left(-\frac{t_\ell-t_0}\tau\right)~~;
\end{lefteqnarray}
where the special case, $t-t_0=t_{1/2}$, yields $1/2=\exp(-t_{1/2}/\tau)$, or:
\begin{lefteqnarray}
\label{eq:tita}
&& t_{1/2}=\tau\ln2~~;
\end{lefteqnarray}
which implies $t_{1/2}<\tau$.

In the case under discussion, $p(\Delta t)=\Delta t/\tau$ via
Eq.\,(\ref{eq:taua}), where $\Delta t\le\tau$ to preserve the statistical
meaning of $p$.   Accordingly, $\Delta t\ll\tau$ implies $p=p(\Delta t)\ll1$
i.e. the validity of the exponential approximation.

The mean lifetime, $\tau$, can be inferred from the intersection of related
curve with the horizontal line, $n^\ast(t)/n_0=1/{\rm e}$, as shown in
Fig.\,\ref{f:cute}.  The same holds for
the mean half-life time, $t_{1/2}$, with regard to the horizontal line,
$n^\ast(t)/n_0=1/2$, as shown in Fig.\,\ref{f:cute}.

The extension of the above definitions to the general case, via
Eqs.\,(\ref{eq:enlr})-(\ref{eq:l}) yields:
\begin{equation}
\label{eq:taur}
\tau=-\frac{\Delta t}{\ln(1-p)}~~;
\end{equation}
and the substitution of Eq.\,(\ref{eq:taur}) into (\ref{eq:enlr}) yields
(\ref{eq:ntaa}).   In the limit, $p\ll1$, $-\ln(1-p)\approx p$ via
Eq.\,(\ref{eq:lose}) and Eq.\,(\ref{eq:taur}) reduces to (\ref{eq:taua}).

In conclusion, Eq.\,(\ref{eq:ntaa}) holds in general for evolving statistical
systems, provided the mean lifetime is defined by Eq.\,(\ref{eq:taur}) instead
of (\ref{eq:taua}).   The same holds for the mean half-life time via
Eq.\,(\ref{eq:tita}), which follows from (\ref{eq:ntaa}).

Values of the mean lifetime, $\tau$, and mean half-life time, $t_{1/2}$, are
listed in Table \ref{t:vime} for the general case, Eq.\,(\ref{eq:taur}), and
the exponential approximation, Eq.\,(\ref{eq:taua}).   An inspection of Table
\ref{t:vime} shows the exponential approximation holds to a good extent in the
limit, $p\ll1$, while the contrary holds for $p\appleq1$, as expected.
\begin{table}
\caption{Fractional mean lifetime, $\tau/\Delta t$, and fractional
half-life time, $t_{1/2}/\Delta t$, as a function of the probability,
$p=p(\Delta t)$, for the general case and the exponential approximation,
indexed as ``gen'' and ``app'', respectively.   See text for further details.}
\label{t:vime}
\begin{center}
\begin{tabular}{lllll}           
\hline
\multicolumn{1}{c}{$p$} &
\multicolumn{1}{c}{$\tau_{\rm gen}/\Delta t$} &
\multicolumn{1}{c}{$\tau_{\rm app}/\Delta t$} &
\multicolumn{1}{c}{$(t_{1/2})_{\rm gen}/\Delta t$} &
\multicolumn{1}{c}{$(t_{1/2})_{\rm app}/\Delta t$} \\
\hline
0.00 & $\infty$     & $\infty$   & $\infty$     & $\infty$        \\
0.01 & 9.9499D$+$01 & 1.0000D+02 & 6.8968D$+$01 & 6.9315D$+$01 \\
0.02 & 4.9498D$+$01 & 5.0000D+01 & 3.4310D$+$01 & 3.4657D$+$01 \\
0.03 & 3.2831D$+$01 & 3.3333D+01 & 2.2757D$+$01 & 2.3105D$+$01 \\
0.04 & 2.4497D$+$01 & 2.5000D+01 & 1.6980D$+$01 & 1.7329D$+$01 \\
0.05 & 1.9496D$+$01 & 2.0000D+01 & 1.3513D$+$01 & 1.3863D$+$01 \\
0.06 & 1.6162D$+$01 & 1.6667D+01 & 1.1202D$+$01 & 1.1552D$+$01 \\
0.07 & 1.3780D$+$01 & 1.4286D+01 & 9.5513D$+$00 & 9.9021D$+$00 \\
0.08 & 1.1993D$+$01 & 1.2500D+01 & 8.3130D$+$00 & 8.6643D$+$00 \\
0.09 & 1.0603D$+$01 & 1.1111D+01 & 7.3496D$+$00 & 7.7016D$+$00 \\
0.10 & 9.4912D$+$00 & 1.0000D+01 & 6.5788D$+$00 & 6.9315D$+$00 \\
0.15 & 6.1531D$+$00 & 6.6667D+00 & 4.2650D$+$00 & 4.6210D$+$00 \\
0.20 & 4.4814D$+$00 & 5.0000D+00 & 3.1063D$+$00 & 3.4657D$+$00 \\
0.30 & 2.8037D$+$00 & 3.3333D+00 & 1.9434D$+$00 & 2.3105D$+$00 \\
0.40 & 1.9576D$+$00 & 2.5000D+00 & 1.3569D$+$00 & 1.7329D$+$00 \\
0.50 & 1.4427D$+$00 & 2.0000D+00 & 1.0000D$+$00 & 1.3863D$+$00 \\
0.60 & 1.0914D$+$00 & 1.6667D+00 & 7.5647D$-$01 & 1.1552D$+$00 \\
0.70 & 8.3058D$-$01 & 1.4286D+00 & 5.7572D$-$01 & 9.9021D$-$01 \\
0.80 & 6.2133D$-$01 & 1.2500D+00 & 4.3068D$-$01 & 8.6643D$-$01 \\
0.90 & 4.3429D$-$01 & 1.1111D+00 & 3.0103D$-$01 & 7.7016D$-$01 \\
0.99 & 2.1715D$-$01 & 1.0101D+00 & 1.5052D$-$01 & 7.0015D$-$01 \\
1.00 & 0.0000D$-$01 & 1.0000D+00 & 0.0000D$-$01 & 6.9315D$-$01 \\
\hline
\end{tabular}
\end{center}
\end{table}                       

\subsection{Radioactive decay} \label{rade}

A sample of radioactive nuclides is a special case of evolving statistical
system, where S$_1$ is a single nuclide, A$_1$ is waiting for a time step,
$\Delta t$, E$_1$ is a nuclide after radioactive decay, $\neg$E$_1$ is a
surviving radioactive nuclide, and $p=p(\Delta t)$ is the probability a
radioactive nuclide decays within a time step, $\Delta t$.

With regard to a selected nuclide, the probability of radioactive decay within
a time step, $\Delta t$, can be inferred from the mean lifetime, $\tau$, or
the mean half-life time, $t_{1/2}$, via Eqs.\,(\ref{eq:tita})-(\ref{eq:taur})
as:
\begin{equation}
\label{eq:radp}
p=1-\exp\left(-\frac{\Delta t}\tau\right)=1-\exp_2\left(-\frac{\Delta t}
{t_{1/2}}\right)~~;
\end{equation}
where $\exp_a(x)=a^x$ and $\exp_e(x)=\exp(x)=e^x$ according to the standard
notation.

Then $p$ depends on the ratio, $\Delta t/\tau$ or $\Delta t/t_{1/2}$: for
instance, $p=9.995\,10^{-4}$ relates to $\Delta t/t_{1/2}=10^{-3}$, regardless
the sample of radioactive nuclides is made of, say, neutrons ($^0$n,
$t_{1/2}=10.183{\rm m}=1.9374\cdot10^{-5}{\rm y}$ \cite{nd16}), or cobalt
($^{60}$Co, $t_{1/2}=5.2747{\rm y}$ \cite{nd16}), or uranium ($^{238}$U,
$t_{1/2}=4.468\cdot10^9{\rm y}$ \cite{nd16}), implying different time steps,
$\Delta t$.   In other words, radioactive nuclides with equal $\Delta t/\tau$
exhibit same $p=p(\Delta t)$.

For extensive results on nuclide mean half-life times and related theory, an
interested reader is addressed to current data collections e.g., \cite{son08}
\cite{ts10} \cite{pst14} and specific textbooks e.g., \cite{del10}
\cite{ben13}, respectively.

\subsection{Passed exam} \label{paex}

An academic course is a special case of evolving statistical system, where
S$_1$ is a single student, A$_1$ is trying an exam, E$_1$ is a student who has
passed an exam, $\neg$E$_1$ is a student who unsucceded in passing an exam,
and $p=p(\Delta t)$ is the probability of passing an exam within a time step,
$\Delta t$.   Clearly an academic course is largely less populated than a
sample of radioactive nuclides, which implies considerable fluctuations with
respect to the expected evolution, as shown in Fig.\,\ref{f:nebi} for
$n_0=10$.   Nevertheless, the mean lifetime and the mean half-life time of an
academic course can be defined as in the case of radioactive decay.

\section{Application to academic courses} \label{aaco}

Avaliable data concern experimentation-of-physic (EOP) courses performed
every academic year (AY) within the range, 1979/80-1998/99, with two
additional occurrences related to 2005/06 and 20013/14, respectively.
Different periods, namely 1979/80-1992/93, 1993/94-1998/99, and 2005/06 +
2013/14, obey different guidelines.

Students are admitted to exam after a selection, where a negative response
implied attendance at course on the next AY.   Students selected as
suitable for exam are considered for random evolution of their course. Related
numbers, $n_0$, for each AY, are listed in Table \ref{t:reco} together with
additional data: for complete explanation, an interested reader is addressed
to Appendix \ref{a:daco}.
\begin{table}
\caption{With regard to a selected experimentation-of-physics (EOP) course,
the following number of graduate
students are listed per academic year: first registation, 
$N_I$; additional registration: unsuitable for exam, $N_R$,
suitable for exam, $N_{R^\ast}$, total, $N_I+N_R+N_R^\ast=N_T$; suitable for
exam at the end of course: first registation, $N_{IP}$,
additional registration, $N_{RP}$, total, 
$N_{IP}+N_{RP}=N_{TP}$; suitable for exam at the end of course but transferred
elsewhere: first registation, $N_{IF}$, additional registration,
$N_{RF}$, total, $N_{IF}+N_{RF}=N_{TF}$; unsuitable for exam at
the end of course: first registation, $N_{IN}$, additional registration,
$N_{RN}$, total, $N_{IN}+N_{RN}=N_{TN}$; inferred suitable 
impostors, $N_{SI}$.   By definition, $N_T=N_{TP}+N_{TF}+N_{TN}$.   Blank 
boxes correspond to lack of data.   To save space, academic years are
labelled by the last two digits and number  columns by related subscripts in
small letters.   Courses obeying different guidelines are
subgrouped by horizontal lines.  See text for further details.}
\label{t:reco}
\begin{center}
\begin{tabular}{|c|r|r|r|r|r|r|r|r|r|r|r|r|r|r|}
\hline
\multicolumn{1}{c}{a.year} &
\multicolumn{1}{c}{i} &
\multicolumn{1}{c}{r} &
\multicolumn{1}{c}{r$^\ast$} &
\multicolumn{1}{c}{t} &
\multicolumn{1}{c}{ip} &
\multicolumn{1}{c}{rp} &
\multicolumn{1}{c}{tp} &
\multicolumn{1}{c}{if} &
\multicolumn{1}{c}{rf} &
\multicolumn{1}{c}{tf} &
\multicolumn{1}{c}{in} &
\multicolumn{1}{c}{rn} &
\multicolumn{1}{c}{tn} &
\multicolumn{1}{c}{si} \\
\hline
79/80 & 38 &    &    &  38 & 30 &    & 30 & 0 &   & 0 &  8 &   &  8 & 22 \\
80/81 & 23 &  1 &  0 &  24 & 12 &  1 & 13 & 2 & 0 & 2 &  9 & 0 &  9 &  4 \\
81/82 & 31 &  2 &  0 &  33 & 23 &  0 & 23 & 1 & 0 & 1 &  7 & 2 &  9 &  7 \\
82/83 & 26 &  2 &  0 &  28 & 20 &  2 & 22 & 0 & 0 & 0 &  6 & 0 &  6 &  9 \\
83/84 & 35 &  3 &  0 &  38 & 34 &  3 & 37 & 0 & 0 & 0 &  1 & 0 &  1 & 23 \\
84/85 & 38 &  3 &  0 &  41 & 30 &  2 & 32 & 0 & 0 & 0 &  8 & 1 &  9 & 13 \\
85/86 & 46 &  1 &  0 &  47 & 45 &  1 & 46 & 1 & 0 & 1 &  0 & 0 &  0 & 27 \\
86/87 & 41 &  5 &  0 &  46 & 35 &  5 & 40 & 1 & 0 & 1 &  5 & 0 &  5 & 16 \\
87/88 & 40 &  5 &  0 &  45 & 33 &  5 & 38 & 0 & 0 & 0 &  7 & 0 &  7 & 23 \\
88/89 & 48 &  7 &  0 &  55 & 44 &  7 & 51 & 0 & 0 & 0 &  4 & 0 &  4 & 32 \\
89/90 & 56 &  1 &  0 &  57 & 45 &  1 & 46 & 0 & 0 & 0 & 11 & 0 & 11 & 28 \\
90/91 & 50 &  7 &  0 &  57 & 42 &  7 & 49 & 0 & 0 & 0 &  8 & 0 &  8 & 29 \\
91/92 & 38 &  3 &  0 &  41 & 32 &  3 & 35 & 0 & 0 & 0 &  6 & 0 &  6 & 16 \\
92/93 & 44 &  1 &  0 &  45 & 30 &  1 & 31 & 0 & 0 & 0 & 14 & 0 & 14 & 18 \\
\hline
93/94 & 93 &  9 &  0 & 102 & 79 &  9 & 88 & 0 & 0 & 0 & 14 & 0 & 14 & 51 \\
94/95 & 64 &  2 & 22 &  88 & 39 &  2 & 41 & 0 & 0 & 0 & 25 & 0 & 25 & 25 \\
95/96 & 71 & 12 & 29 & 112 & 30 & 12 & 42 & 0 & 0 & 0 & 41 & 0 & 41 & 29 \\
96/97 & 20 & 15 & 13 &  48 & 15 & 15 & 30 & 0 & 0 & 0 &  5 & 0 &  5 & 20 \\
97/98 & 64 &  2 &  5 &  71 & 53 &  2 & 55 & 0 & 0 & 0 & 11 & 0 & 11 & 49 \\
98/99 & 56 &  3 &  7 &  66 & 48 &  2 & 50 & 0 & 0 & 0 &  8 & 1 &  9 & 47 \\
\hline
05/06 & 37 &    &    &  37 & 36 &    & 36 & 0 &   & 0 &  1 &   &  1 &  9 \\
13/14 & 46 &    &    &  46 & 39 &    & 39 & 0 &   & 0 &  7 &   &  7 & 17 \\
\hline
\end{tabular}                 
\end{center}                      
\end{table}                       

\begin{figure*}[t]  
\begin{center}      
\includegraphics[scale=0.8]{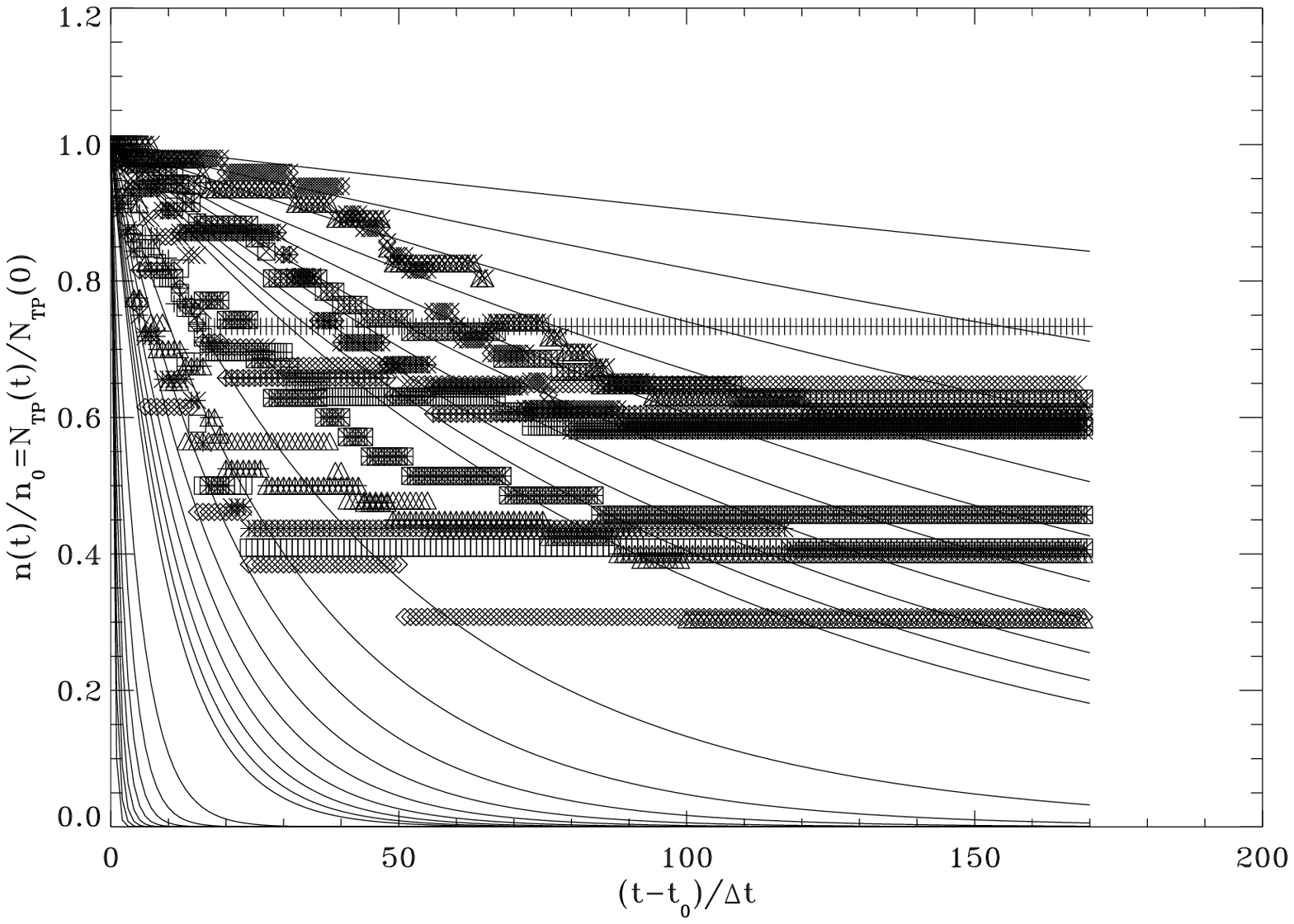}                      
\caption[ddbb]{Empirical random evolution of the fractional number of
surviving units,
$n(t)/n_0=N_{TP}(t)/N_{TP}(0)$, inferred from data collections related to
experimentation-of-physics (EOP) courses per selected academic year (AY), as
listed in
Table \ref{t:reco}.   The number of suitable impostors is understimated as
$N_{SI}=0$.   The time step is $\Delta t=(1/12)$y.   Symbol caption per AY:
1979/80 - crosses; 1980/81 - diamonds; 1981/82 - triangles; 1982/83 - squares;
1983/84 - saltires; 1984/85 - asterisks; 1985/86 - crosses \& squares; 1986/87
- crosses \& triangles; 1987/88 - crosses \& diamonds; 1988/89 - saltires \&
squares; 1989/90 - saltires \& triangles; 1990/91 - saltires \& diamonds;
1991/92 - asterisks \& squares; 1992/93 - asterisks \& diamonds; where ``\&''
has to be read as ``superimposed to''.   Curves represent expected evolution
related to different probabilities, $p=\exp_{10}(-k)i$, $1\le i\le9$,
$1\le k\le3$, where larger values denote lower curves and vice versa.
See text for further details.}
\label{f:corg2}     
\end{center}       
\end{figure*}

\begin{figure*}[t]  
\begin{center}      
\includegraphics[scale=0.8]{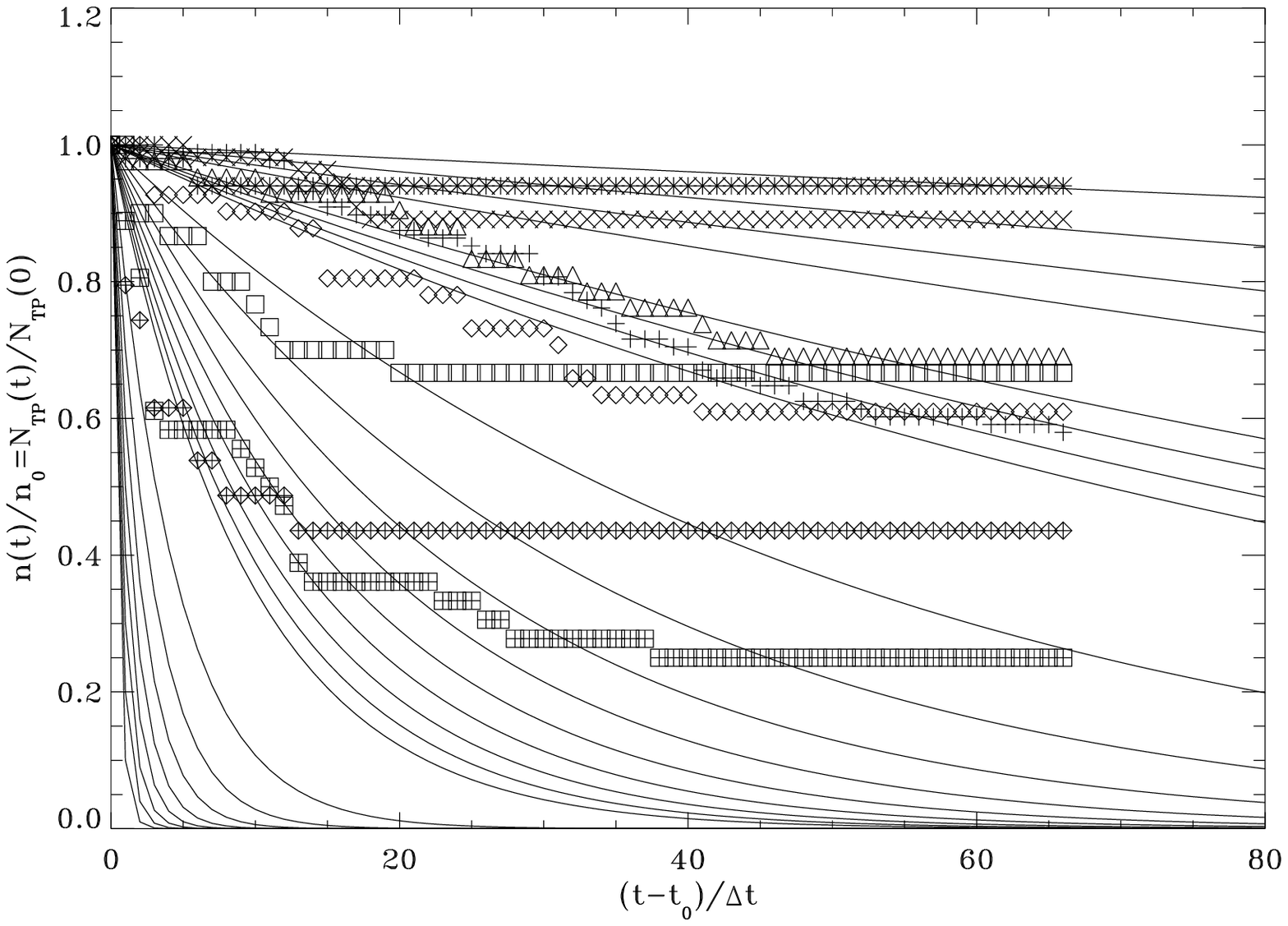}                      
\caption[ddbb]{Empirical random evolution of the fractional number of
surviving units,
$n(t)/n_0=N_{TP}(t)/N_{TP}(0)$, inferred from data collections related to
experimentation-of-physics (EOP) courses per selected academic year (AY), as
listed in
Table \ref{t:reco}.   The number of suitable impostors is understimated as
$N_{SI}=0$.   The time step is $\Delta t=(1/12)$y.   Symbol caption per AY:
1993/94 - crosses; 1994/95 - diamonds; 1995/96 - triangles; 1996/97 - squares;
1997/98 - saltires; 1998/99 - asterisks; 2005/06 - crosses \& squares; 2013/14
- crosses \& diamonds; where ``\&''
has to be read as ``superimposed to''.   Curves represent expected evolution
related to different probabilities, $p=\exp_{10}(-k)i$, $1\le i\le9$,
$1\le k\le3$, where larger values denote lower curves and vice versa.
See text for further details.}
\label{f:corg1}     
\end{center}       
\end{figure*}

Empirical random evolutions related to data collections are plotted in
Figs.\,\ref{f:corg2} and \ref{f:corg1} for AY within the range,
1979/80-1992/93 and 1993/94-1998/99 + 2005/06 + 2013/14, respectively.   Also
shown therein are expected evolutions related to $p=10^{-k}i$, $1\le i\le9$,
$1\le k\le3$, where curves from up to down correspond to increasing $p$.
An inspection of Figs.\,\ref{f:corg2} and \ref{f:corg1} shows empirical random
evolutions are affected by considerable statistical fluctuations, as expected
from the low population of related samples.   In addition, empirical random
evolutions never exceed a threshold above zero, which is at odds with an
assumed absence of students who transferred elsewhere or abandoned university,
or ``suitable impostors'' (SI), $N_{SI}=0$.   For further details, an
interested reader is addressed to Appendix \ref{a:daco}.

\begin{figure*}[t]  
\begin{center}      
\includegraphics[scale=0.8]{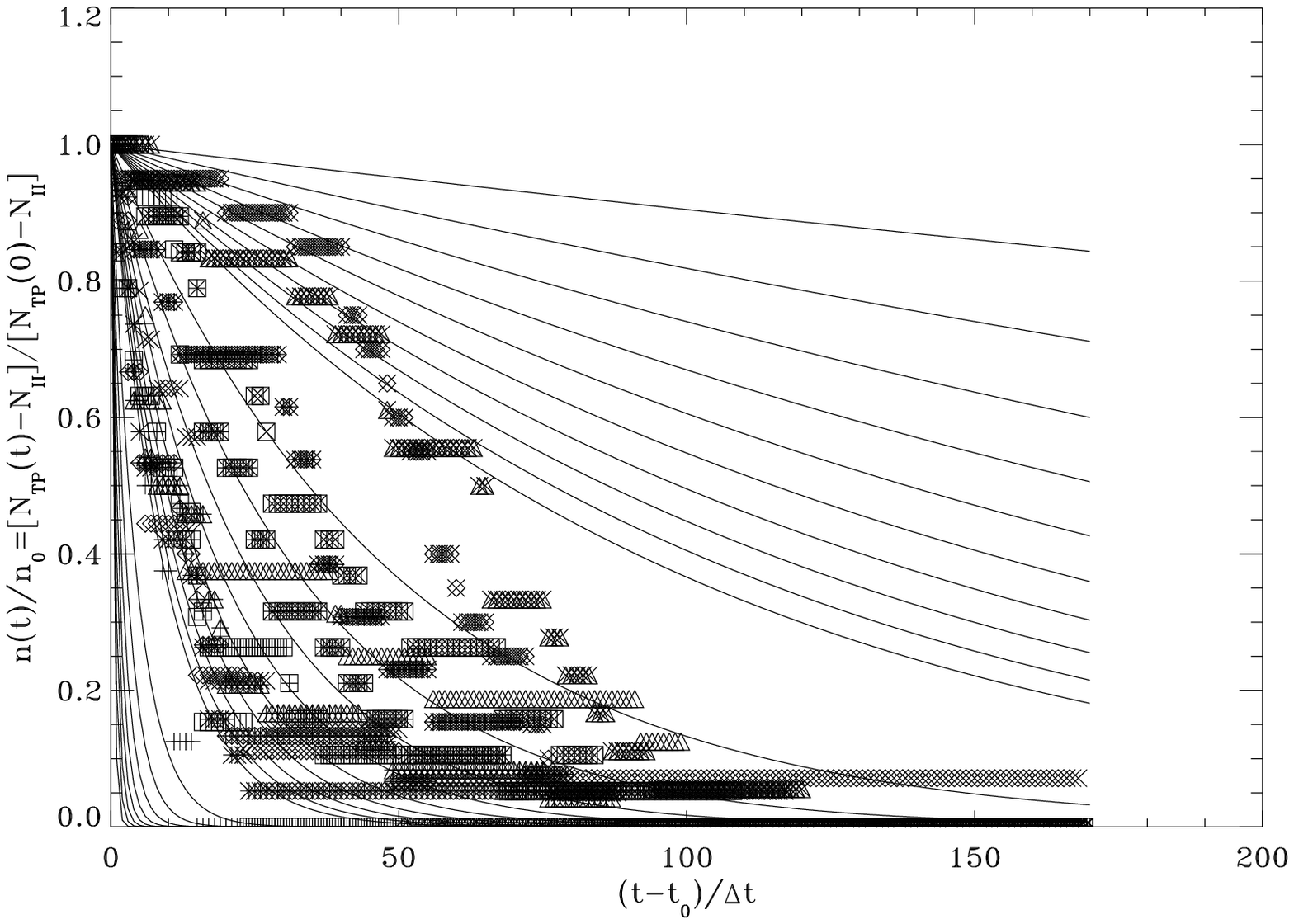}                      
\caption[ddbb]{Random evolution of the fractional number of surviving units,
$n(t)/n_0=[N_{TP}(t)-N_{TP}(t_L)]/[N_{TP}(0)-N_{TP}(t_L)]$, inferred from data
collections related to experimentation-of-physics (EOP) courses per selected
academic year (AY), as listed in Table \ref{t:reco}.   The number of suitable
impostors is overstimated as $N_{SI}=N_{TP}(t_L)$.   The time step is
$\Delta t=(1/12)$y.   Symbol caption and curves as in Fig.\,\ref{f:corg2}.
See text for further details.}
\label{f:corg4}     
\end{center}       
\end{figure*}

\begin{figure*}[t]  
\begin{center}      
\includegraphics[scale=0.8]{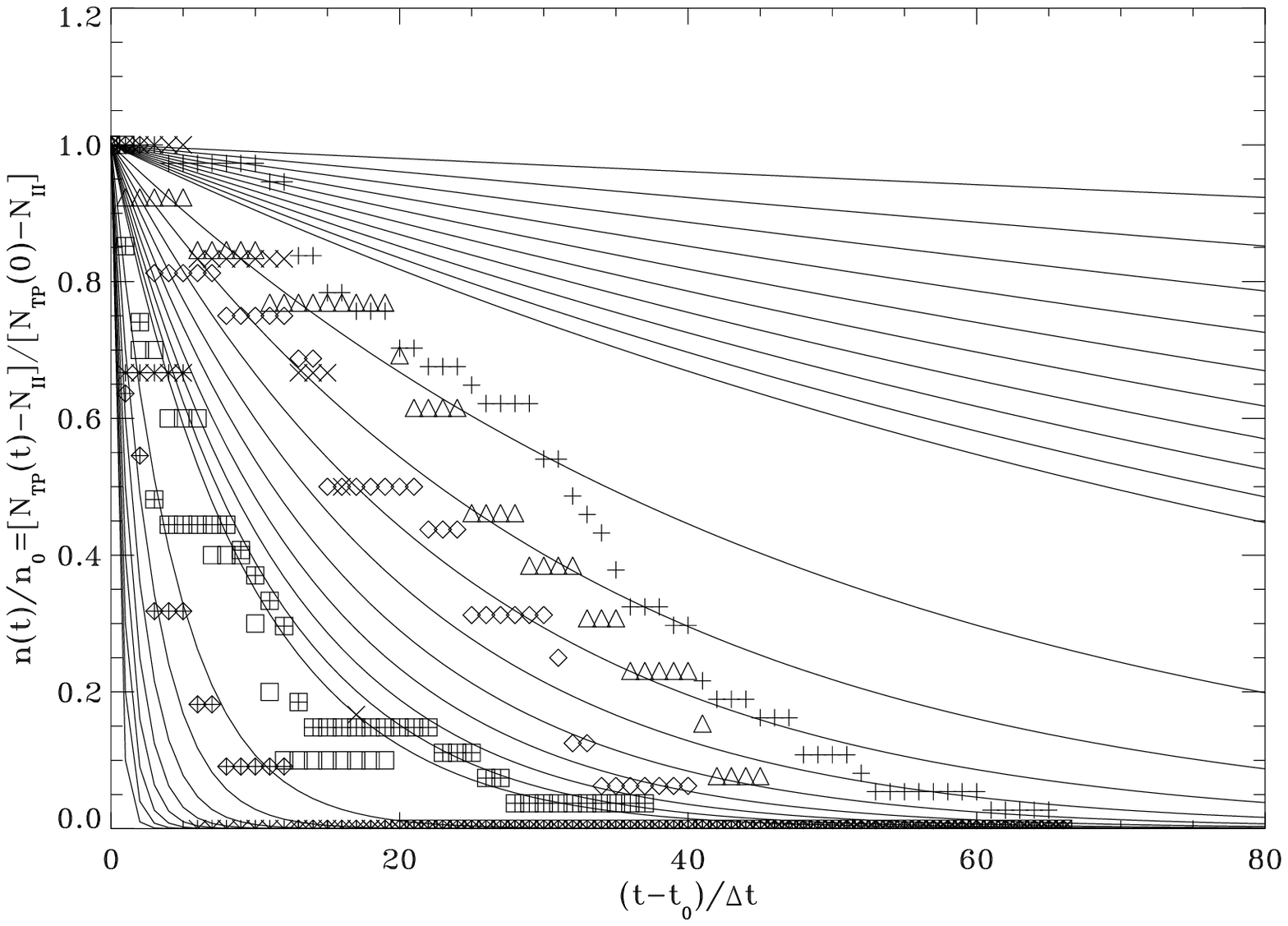}                      
\caption[ddbb]{Empirical random evolution of the fractional number of
surviving units,
$n(t)/n_0=[N_{TP}(t)-N_{TP}(t_L)]/[N_{TP}(0)-N_{TP}(t_L)]$, inferred from data
collections related to experimentation-of-physics (EOP) courses per selected
academic year (AY), as listed in Table \ref{t:reco}.   The number of suitable
impostors is overstimated as $N_{SI}=N_{TP}(t_L)$.   The time step is
$\Delta t=(1/12)$y.   Symbol caption and curves as in Fig.\,\ref{f:corg1}.
See text for further details.}
\label{f:corg3}     
\end{center}       
\end{figure*}

Taking an upper value, $N_{SI}=N_{TP}(t_L)$, where $t_L$ is the ending time
related to data collection, yields empirical random evolutions plotted in 
Figs.\,\ref{f:corg4} and \ref{f:corg3}, respectively, while expected
evolutions remain unchanged.   
An inspection of Figs.\,\ref{f:corg4} and \ref{f:corg3} shows empirical random
evolutions decline to a comparable extent with respect to expected
evolutions. Statistical fluctuations look similar to their counterparts
exhibited by model random evolutions inferred from a sequence of random
numbers in connection with low-population samples, as depicted in
Fig.\,\ref{f:nebi}.

\begin{figure*}[t]  
\begin{center}      
\includegraphics[scale=0.8]{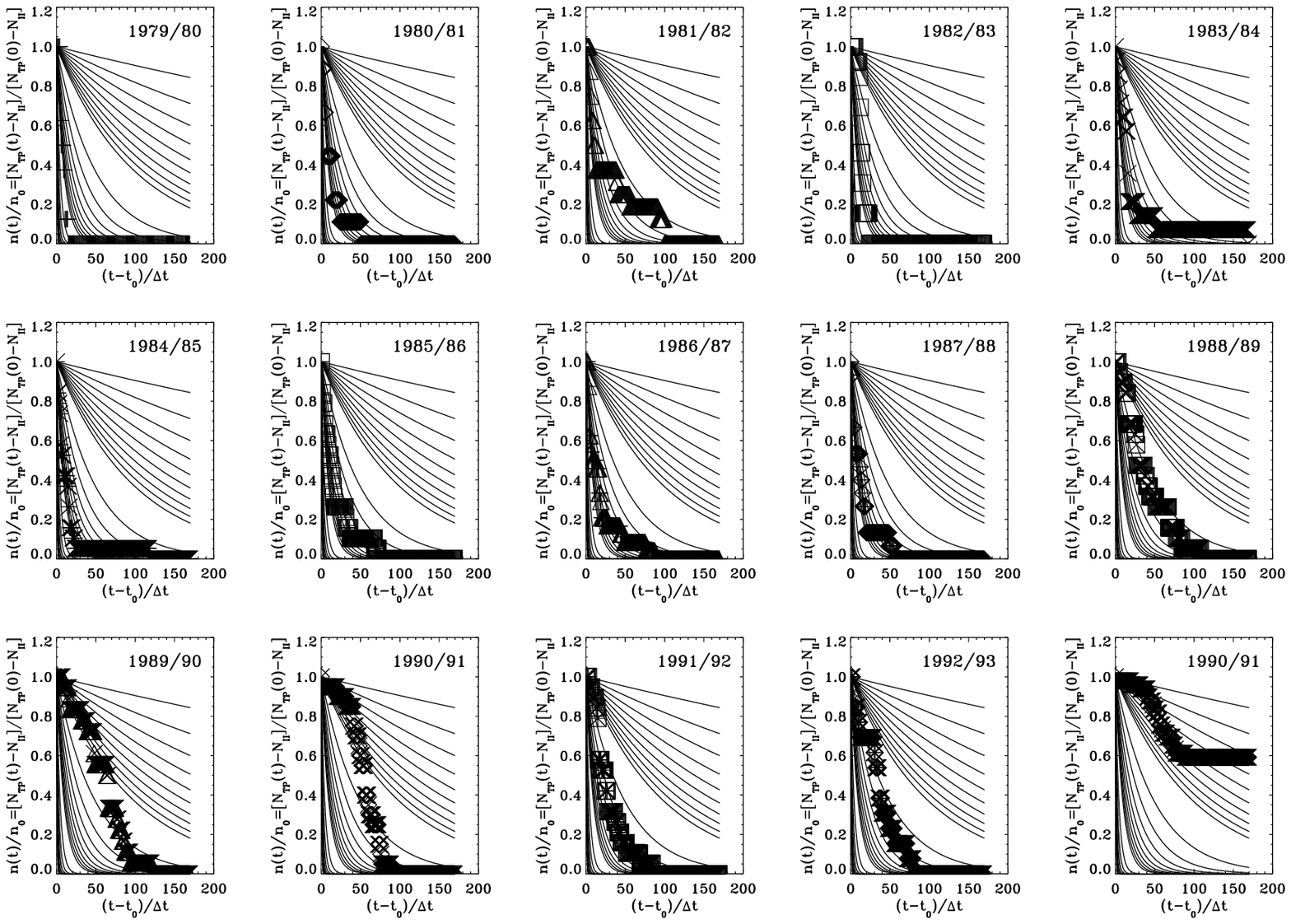}                      
\caption[ddbb]{Empirical random evolution shown in Fig.\,\ref{f:corg4},
plotted on
different panels for different academic years (top labels).   Curves as in
Fig.\,\ref{f:corg4}.  The bottom right panel, related to Fig.\,\ref{f:corg2},
is placed for comparison.   See text for further details.}
\label{f:cg415}     
\end{center}       
\end{figure*}

\begin{figure*}[t]  
\begin{center}      
\includegraphics[scale=0.8]{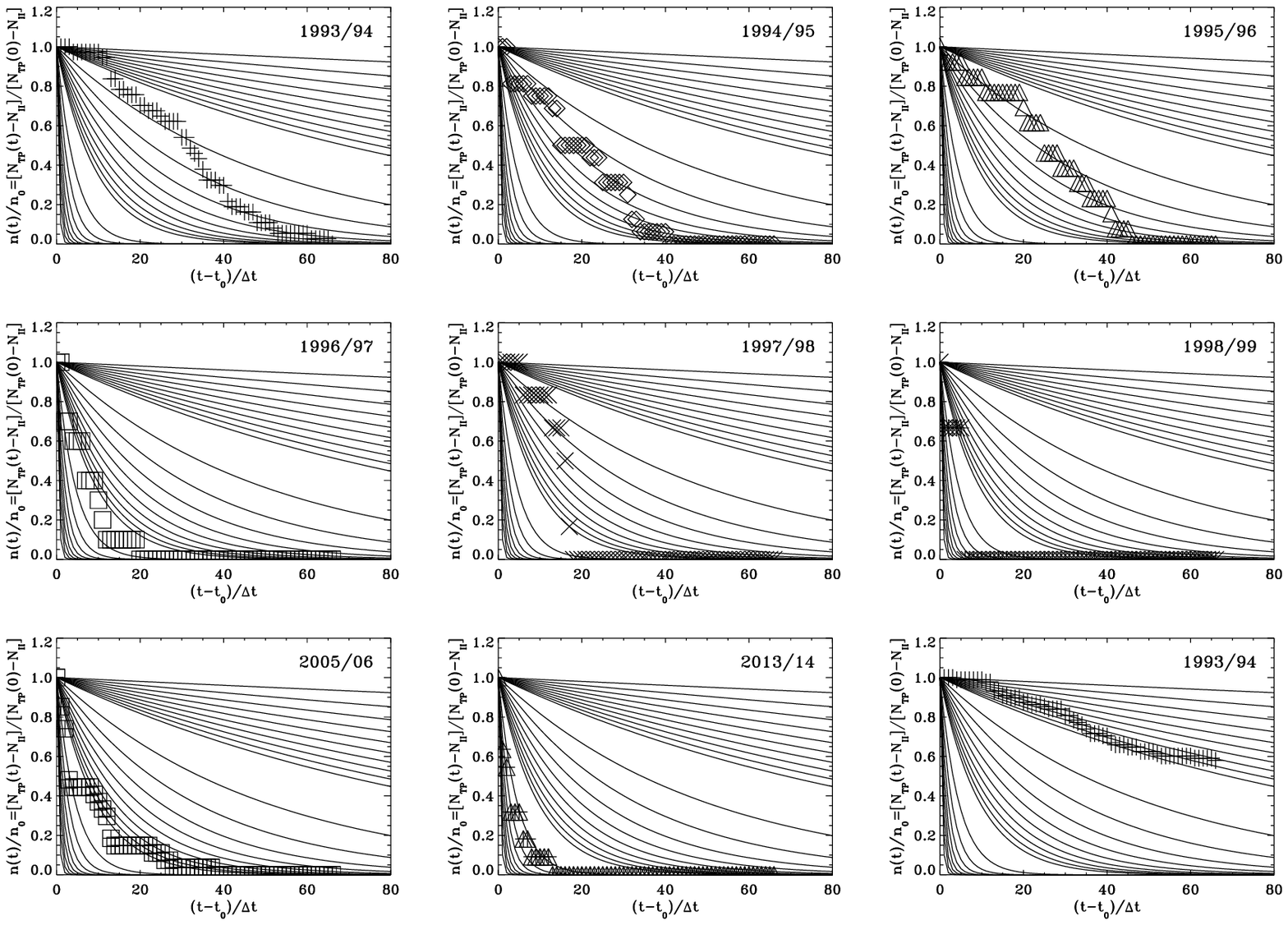}                      
\caption[ddbb]{Empirical random evolution shown in Fig.\,\ref{f:corg3},
plotted on
different panels for different academic years (top labels).   Curves as in
Fig.\,\ref{f:corg3}.  The bottom right panel, related to Fig.\,\ref{f:corg1},
is placed for comparison.   See text for further details.}
\label{f:cg308}     
\end{center}       
\end{figure*}

Empirical random evolutions, shown collectively in Figs.\,\ref{f:corg4} and
\ref{f:corg3}, are plotted separately in Figs.\,\ref{f:cg415} and
\ref{f:cg308}, respectively, where related AY is labelled on each panel.
An inspection of Figs.\,\ref{f:cg415} and \ref{f:cg308} shows empirical random
evolutions fit to expected evolutions to a different extent, ranging between
close agreement e.g., AY 1984/85 and marked disagreement e.g., AY 1990/91.
The bottom right panel in Figs.\,\ref{f:cg415} and \ref{f:cg308} relates to
Figs.\,\ref{f:corg4} and \ref{f:corg3}, respectively, where SI number has been
assumed equal to zero, and is placed for comparison.

\begin{figure*}[t]  
\begin{center}      
\includegraphics[scale=0.8]{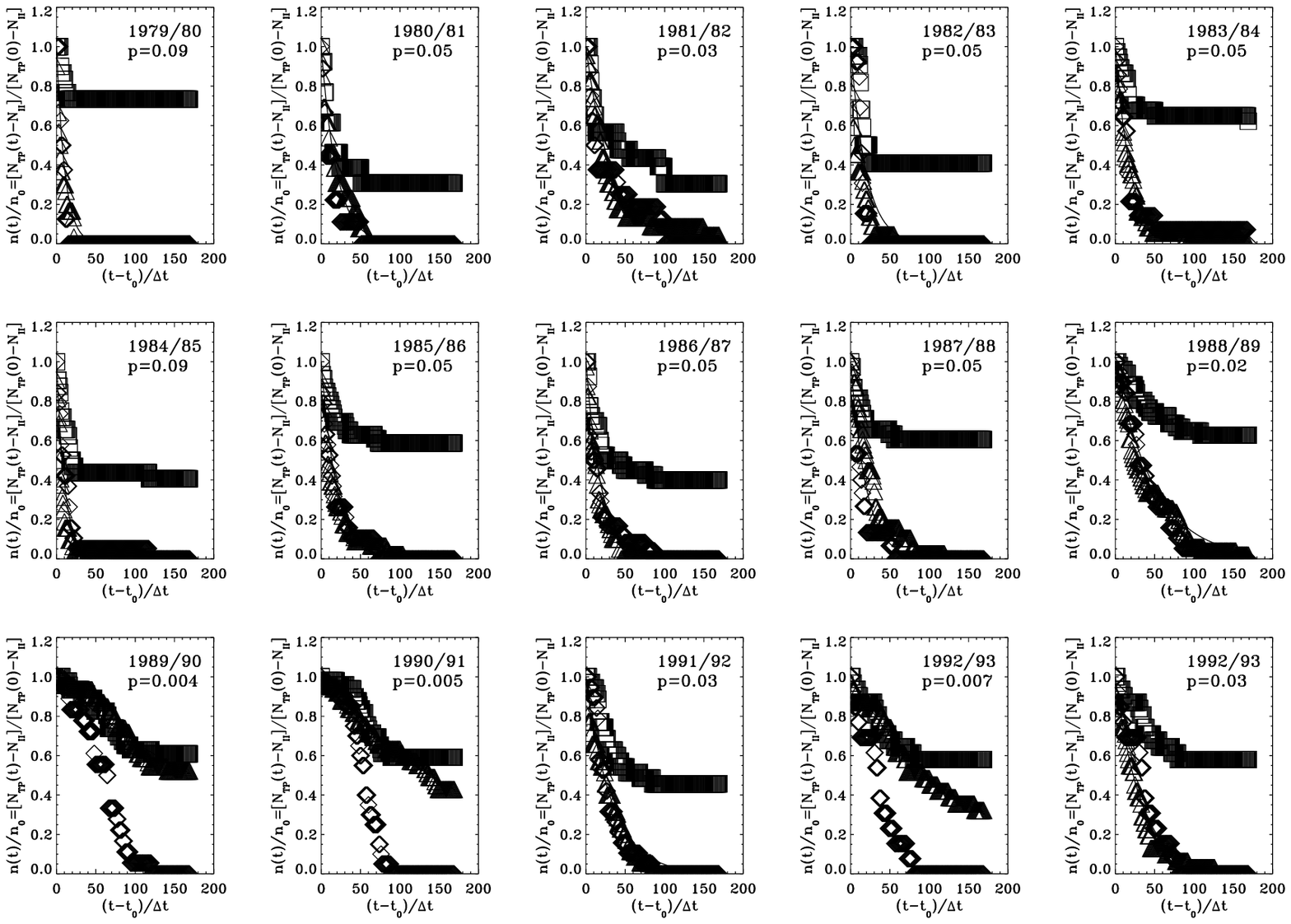}                      
\caption[ddbb]{Empirical random evolution shown in Fig.\,\ref{f:corg2}
(squares) and in Fig.\,\ref{f:corg4} (diamonds), plotted on
different panels for different academic years (top labels); expected evolution
(curve) for a selected value of the probability, $p$ (bottom labels); related
model random evolution (triangles) inferred from a sequence of random numbers.
The bottom right panel is a repetition of related empirical random evolution,
but with different expected and model random evolution plotted therein.   See
text for further details.}
\label{f:rg415}     
\end{center}       
\end{figure*}

\begin{figure*}[t]  
\begin{center}      
\includegraphics[scale=0.8]{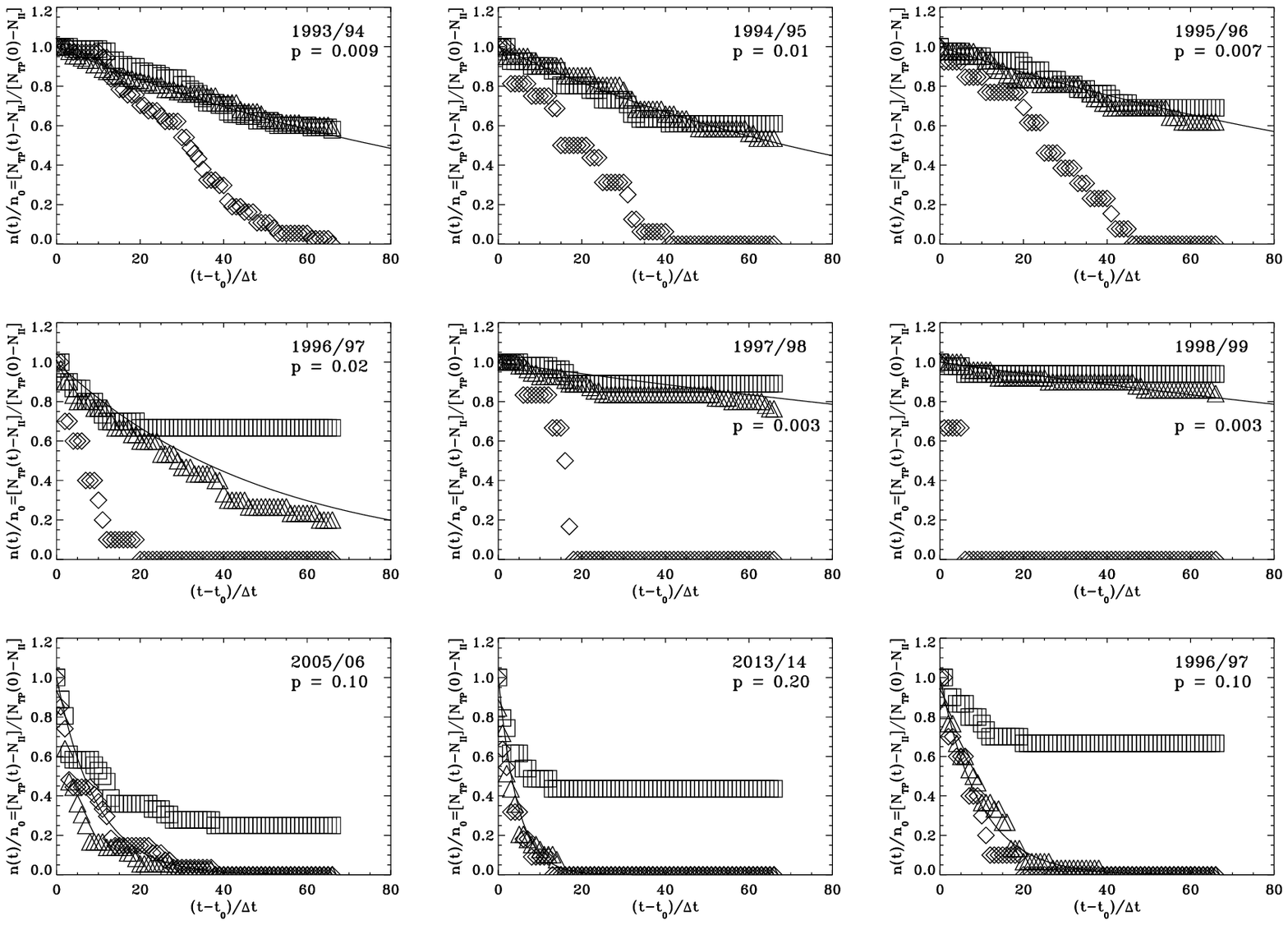}                      
\caption[ddbb]{Empirical random evolution shown in Fig.\,\ref{f:corg1}
(squares) and in Fig.\,\ref{f:corg3} (diamonds), plotted on
different panels for different academic years (top labels); expected evolution
(curve) for a selected value of the probability, $p$ (bottom labels); related
model random evolution (triangles) inferred from a sequence of random numbers.
The bottom right panel is a repetition of related empirical random evolution,
but with different expected and model random evolution plotted therein.   See
text for further details.}
\label{f:rg308}     
\end{center}       
\end{figure*}

The effect of statistical fluctuations is outlined in Figs.\,\ref{f:rg415} and
\ref{f:rg308}, where empirical random evolutions are as in
Figs.\,\ref{f:corg2} (squares) - \ref{f:corg4} (diamonds) and in
Figs.\,\ref{f:corg1} (squares) - \ref{f:corg3} (diamonds), respectively, with
the addition of a single expected evolution (curve) related to an assigned
probability, $p$, and a model random evolution (triangles) inferred from a
sequence of random numbers, as in Fig.\,\ref{f:nebi}.
An inspection of
Figs.\,\ref{f:rg415} and \ref{f:rg308} discloses empirical random evolutions
are consistent with related expected evolutions, in that statistical
fluctuations appear to be of comparable order with respect to their
counterparts exhibited by model random evolutions.

For data collections spanning over sufficiently large time intervals e.g., AY
1979/80-1988/89, expected evolutions are closer to empirical random evolutions
where an upper limit to SI number is assumed; for sufficiently short time
intervals e.g., AY 1993/94-1998/99, expected evolutions are closer to
empirical random evolutions where a lower limit (equal to zero) to SI number
is assumed; for intermediate time intervals e.g., AY 1989/90-1992/93, expected
evolutions are close to empirical random evolutions regardless of assumed SI
number; for the shortest time intervals e.g., AY 2005/06 and 2013/14, expected
evolutions are closer again to empirical random evolutions where an upper
limit to SI number is assumed.

In conclusion, empirical random distributions, related to EOP courses under
consideration, can safely be described via the binomial distribution, as
outlined in Section \ref{ests}, where probabilities lie within the range,
$0.003\le p\le0.200$, with regard to a time step, $\Delta t=(1/12)$y.

\section{Discussion}
\label{disc}

Evolving statistical systems, where only a possible event and the opposite
event are involved, have a wide range of applications in spite of their
intrinsic simplicity, in particular radioactive decay for high-population
samples $(N\gg10)$ and academic courses for low-population samples
$(N\appgeq10)$.

The expected evolution of the fractional number of surviving units,
$n_\ell^\ast/n_0$, is expressed by Eq.\,(\ref{eq:ntaa}) in the limit, $p\ll1$,
if the mean lifetime is expressed as $\tau=\Delta t/p$, where $p$ is the
probability of the possible event during a time step,
$\Delta t=t_k-t_{k-1}$.   On the other hand, Eq.\,(\ref{eq:ntaa}) is exact
provided the mean lifetime is defined as $\tau=-\Delta t/\ln(1-p)$.   It is
worth emphasizing $p$ depends on the ratio, $\tau/\Delta t$, as shown in Table
\ref{t:vime}, or in other words expected statistical evolution can be scaled
to the ratio, $\tau/\Delta t$.

Concerning the application to passed exams, the key factors for interpreting
academic courses as evolving statistical systems, where $p$ is the probability
of passing an exam within a time step, $\Delta t$, are essentially two, namely
estimation of SI number, $N_{SI}$, and role of statistical fluctuations.

A lower limit to $N_{SI}$ is clearly zero, while an upper limit can be
inferred from a flat tail shown by empirical random evolution,
$n_\ell/n_0$, for sufficiently long times.   The true evolution lies between
the above mentioned extreme cases.    SI take origin from several independent
occurrences, such as incomplete knowledge on students who decided to transfer
or dismiss or suspend academic studies, lack of data on passed exams, and so
on.

An upper limit to $N_{SI}$ can safely be assumed for data collections
spanning over a wide time range, while a lower limit could be preferred in
connection with narrower intervals, as depicted in Figs.\,\ref{f:rg415} and
\ref{f:rg308}.   Displacements of empirical random evolutions from related
expected evolutions are comparable to their counterparts related to model
random evolutions inferred from sequences of random numbers, as shown in
Figs.\,\ref{f:rg415} and \ref{f:rg308}.

In addition, empirical random evolutions are weakly dependent on course
guidelines, in the sense that discrepancies between empirical random
evolutions related to EOP courses following equal guidelines (AY
1979/80-1992/93; 1993/94-1998/99; 2005/2006 + 2013/14) are comparable to
their counterparts related to EOP courses following different guidelines, as
depicted in Figs.\,\ref{f:rg415} and \ref{f:rg308}.   Expected evolutions
plotted therein are arbitrarily selected regardless of fitting procedures,
aiming to show they mimic related empirical random evolutions to an acceptable
extent.

Accordingly, the probability of passing a EOP exam within a time
step, $\Delta t=(1/12)$y, may safely be assumed as time independent.   
Inferred mean lifetimes, $\tau$, and half-life times, $t_{1/2}$, are listed in
Table \ref{t:taue}, where larger values relate to data collections spanning
over a short time interval, with the exception of AY 2005/06 and 2013/14 which
were subjected to different guidelines.
\begin{table}
\caption{Mean lifetime, $\tau$, and mean half-life time, $t_{1/2}$, of
experimentation-of-physics (EOP) courses per academic year (AY), inferred from
the probability, $p=p(\Delta t)$, $\Delta t=(1/12)$y, related to expected
evolutions fitting to empirical random evolutions plotted in
Figs.\,\ref{f:rg415}-\ref{f:rg308}.   To save space, AYs are
labelled by the last two digits.   Courses performed following different
guidelines are subgrouped by horizontal lines.}
\label{t:taue}
\begin{center}
\begin{tabular}{llll}           
\hline
\multicolumn{1}{c}{a.year} &
\multicolumn{1}{c}{$p$} &
\multicolumn{1}{c}{$\tau/{\rm y}$} &
\multicolumn{1}{c}{$t_{1/2}/{\rm y}$} \\
\hline
79/80 & 0.09  & 8.83604D$-$1 & 6.12468D$-$1 \\
80/81 & 0.05  & 1.62464D$+$0 & 1.12612D$+$0 \\
81/82 & 0.02  & 4.12486D$+$0 & 2.85913D$+$0 \\
82/83 & 0.05  & 1.62464D$+$0 & 1.12612D$+$0 \\
83/84 & 0.05  & 1.62464D$+$0 & 1.12612D$+$0 \\
84/85 & 0.09  & 8.83604D$-$1 & 6.12468D$-$1 \\
85/86 & 0.05  & 1.62464D$+$0 & 1.12612D$+$0 \\
86/87 & 0.05  & 1.62464D$+$0 & 1.12612D$+$0 \\
87/88 & 0.05  & 1.62464D$+$0 & 1.12612D$+$0 \\
88/89 & 0.02  & 4.12486D$+$0 & 2.85913D$+$0 \\
89/90 & 0.004 & 2.07916D$+$1 & 1.44117D$+$1 \\
90/91 & 0.005 & 1.66250D$+$1 & 1.15235D$+$1 \\
91/92 & 0.03  & 2.73590D$+$0 & 1.89638D$+$0 \\
92/93 & 0.007 & 1.18630D$+$1 & 8.22284D$+$0 \\
\hline                                    
93/94 & 0.009 & 9.21753D$+$0 & 6.38910D$+$0 \\
94/95 & 0.01  & 8.29160D$+$0 & 5.74730D$+$0 \\
95/96 & 0.007 & 1.18630D$+$1 & 8.22284D$+$0 \\
96/97 & 0.02  & 4.12486D$+$0 & 2.85913D$+$0 \\
97/98 & 0.003 & 2.77361D$+$1 & 1.92252D$+$1 \\
98/99 & 0.003 & 2.77361D$+$1 & 1.92252D$+$1 \\
\hline                                    
05/06 & 0.1   & 7.90935D$-$1 & 5.48234D$-$1 \\
13/14 & 0.2   & 3.73452D$-$1 & 2.58857D$-$1 \\   
\hline
\end{tabular}
\end{center}
\end{table}                       
Then long mean lifetimes and half-life times are probably overstimated, due to
incomplete data collections.

\section{Conclusion} \label{conc}

Statistical systems have been conceived from the standpoint of statistical
mechanics, as made of a (generally large) number of identical units and
exhibiting a (generally large) number of different configurations
(microstates), among which only equivalence classes (macrostates) are
accessible to observations.

Further attention has been devoted to evolving
statistical systems and a simple attempt, involving only a possible event and
the
opposite event, is examined in detail.   In particular, the expected evolution
has been determined and compared to the random evolution inferred from a
sequence of random numbers, for different sample populations.

The special case of radiactive decay has been considered and results have been
expressed in terms of fractional time, $t/\Delta t$, where the time step,
$\Delta t$, is related to the decay probability, $p=p(\Delta t)$.

An application to data collection related to experimentation-of-physics (EOP)
courses per academic year (AY) has shown related
empirical random evolution of the fractional number of surviving units,
$n(t)/n_0$, could be biased by lack of data on students who, for some reason,
dismissed the course, defined as suitable impostors (SI).   The extreme cases,
related to a null and an inferred upper limit to SI number, $N_{SI}$, have
been considered.   A comparison has been performed with expected evolutions
and model random evolutions inferred from sequences of random numbers, for
different values of the probability, $p$, of passing a EOP exam.   The main
results are listed below.
\begin{description}
\item[(1)] At least one among the empirical random evolutions, related to a
null and an inferred upper limit to $N_{SI}$, is fitted by an appropriate
expected evolution to an acceptable extent.
\item[(2)] For data collections spanning over a sufficiently wide time range,
a closer agreement between empirical random evolution and expected evolution
takes place assuming the upper limit to SI number, $N_{SI}=N_{TP}(t_L)$.
\item[(3)] For data collections spanning over a sufficiently narrow time
range, a closer agreement between empirical random evolution and expected
evolution takes place assuming the lower limit to SI number, $N_{SI}=0$.   An
exception arises from the shortest intervals (AY 2005/06 and 2013/14),
possibly due to large lack of data.
\item[(4)] For data collections spanning over an intermediate time
range, the agreement between empirical random evolution and expected
evolution is weakly dependent on the assumed $N_{SI}$.
\item[(5)] Empirical random evolutions exhibit weak dependence on course
guidelines, in the sense that discrepancies from AY obeying equal guidelines
are comparable to discrepancies from AY obeying different guidelines.
\item[(6)] Statistical fluctuations exhibited by empirical and model random
evolution, with respect to related expected evolution, are of comparable
order.
\item[(7)] Inferred values of the probability, $p$, related to the time step,
$\Delta t=(1/12)$y, lie within the range, $0.003\le p\le0.200$.
\end{description}

In conclusion, the evolving statistical system made of an academic course is
similar to a poorly populated sample of radioactive nuclides exhibiting equal
values of probability, $p$, and time step, $\Delta t$.

\appendix
\section*{Appendix}

\section{Data collections and input parameters}\label{a:daco}

Data collections relate to 14 experimentation-of-physics (EOP) academic
courses per academic year (AY).   Owing
to different guidelines implying different materials and methods, the sample
can be divided into three subsamples, namely (a) 1979/80-1992/93, (b)
1993/94-1998/99, and (c) 2005/06 + 2013/14.   More specifically, EOP courses
are structured in the following way.   Lessons and experimentations range
along (a) two consecutive AY;  (b) one AY; (c) one half AY.   In any case,
selection is made on attending experimentations: students exhibiting two
absences or less are suitable for exam, while more than two absences implies
students are unsuitable for exam.

EOP courses can be attended by new students (first registration), in number of
$N_I$, and students who already attended (additional registration), in the
last case either unsuitable or suitable for exam, in number of $N_R$ and
$N_{R^\ast}$, respectively, for a total of $N_T=N_I+N_R+N_{R^\ast}$.

At the end of the course, students suitable for exam are in number of
$N_{IP}$ among initial $N_I$ and $N_{RP}$ among initial $N_R$, for a total of
$N_{TP}=N_{IP}+N_{RP}$.   Students among initial $N_{R^\ast}$ are already
suitable for exam and then are not considered to this respect.
Suitable students among initial $N_I$ and $N_R$, who decided to transfer
elsewhere or to abandon university, are counted apart as $N_{IF}$ and
$N_{RF}$, respectively, for a total of $N_{TF}=N_{IF}+N_{RF}$.   Finally,
students unsuitable for exam are in number of $N_{IN}$ among initial $N_I$ and
$N_{RN}$ among initial $N_R$, for a total of $N_{TN}=N_{IN}+N_{RN}$.

Accordingly, the evolving statistical system made of a selected EOP course has
an initial number of units, $n_0=N_{TP}$, which has to be considered as an
upper limit due to lack of information about students who decided to transfer
elsewhere or to abandon university.   Then the above mentioned ``suitable
impostors'' (SI), in number of $N_{SI}$, should be subtracted from suitable
students, yielding $n_0=N_{TP}-N_{SI}$, $0\le N_{SI}\le N_{TR}(t_L)$, where
$t_L$ is the time at the end of data collection.   For
$(t_L-t_0)/\Delta t\gg1$ and $N_{TP}(t_L)\gg1$, $N_{SI}$ may safely be assumed
equal to $N_{TP}(t_L)$.

In conclusion, with regard to an evolving statistical system made of an
academic course, the fractional number of surviving units lies between the
extreme cases:
\begin{lefteqnarray}
\label{eq:nfi}
&& \frac{n(t)}{n_0}=\frac{N_{TP}(t)}{N_{TP}(0)}~~;\quad n_0=N_{TP}(0)~~; \\
\label{eq:nfs}
&& \frac{n(t)}{n_0}=\frac{N_{TP}(t)-N_{TP}(t_L)}{N_{TP}(0)-N_{TP}(t_L)}~~;
\quad n_0=N_{TP}(0)-N_{TP}(t_L)~~;
\end{lefteqnarray}
where the total number of suitable (for exam) students at the end of the
course, $N_{TP}=N_{TP}(0)$, and the upper limit of SI, $N_{SI}=N_{TP}(t_L)$,
are listed in Table \ref{t:reco}.   The time step is fixed as
$\Delta t=(1/12)$y, in the sense that students were allowed to try exam every
month.

In addition to the parameters shown in Table \ref{t:reco}, data collections
include evolution of surviving units, $n(t)=N_{TP}(t)$, which can be
visualized in Figs.\,\ref{f:corg2} and \ref{f:corg1} regardless of the
vertical scale.   Related tables exhibit about one hundred lines per AY, and
for this reason are not presented here.

\end{document}